\documentclass[aps, prb, reprint, aps, twocolumn, 10pt, superscriptaddress]{revtex4-1}

\usepackage{graphicx, amsmath, dsfont, dcolumn, bm, times, color, braket, amssymb}
\usepackage{graphicx}
\usepackage{float}
\usepackage{dcolumn}
\usepackage{xcolor}

\newcommand{\mgvcomment}[1]{}
\newcommand{\snccomment}[1]{}
\usepackage{soul}

\begin{document}

\bibliographystyle{apsrev4-1}

\title{Majorana bound states in nanowire-superconductor hybrid systems in periodic magnetic fields}

\author{Viktoriia Kornich}
\affiliation{Department of Physics, University of Wisconsin-Madison, Madison, Wisconsin 53706, USA}
\affiliation{Kavli Institute of Nanoscience, Delft University of Technology, 2628 CJ Delft, The Netherlands }
\author{Maxim G.\ Vavilov}
\affiliation{Department of Physics, University of Wisconsin-Madison, Madison, Wisconsin 53706, USA}
\author{Mark Friesen}
\affiliation{Department of Physics, University of Wisconsin-Madison, Madison, Wisconsin 53706, USA}
\author{M.\ A.\ Eriksson}
\affiliation{Department of Physics, University of Wisconsin-Madison, Madison, Wisconsin 53706, USA}
\author{S.\ N.\ Coppersmith}
\affiliation{Department of Physics, University of Wisconsin-Madison, Madison, Wisconsin 53706, USA}
\affiliation{School of Physics, The University of New South Wales, Sydney NSW 2052, Australia}

\date{\today}

\begin{abstract}
We study how the shape of a periodic magnetic field affects the presence of  Majorana bound states (MBS) in a nanowire-superconductor system. Motivated by the field configurations that can be produced by an array of nanomagnets, we consider spiral fields with an elliptic cross-section and fields with two sinusoidal components. We show that MBS are robust to imperfect helical magnetic fields.  In particular, if the amplitude of one component is tuned to the value determined by the superconducting order parameter in the wire, the MBS can exist even if the second component has a much smaller amplitude.  We also explore the effect of the chemical potential on the phase diagram.
Our analysis is both numerical and analytical, with good agreement between the two methods.

 \end{abstract}

\maketitle

\let\oldvec\vec
\renewcommand{\vec}[1]{\ensuremath{\boldsymbol{#1}}}

\section{introduction}
\label{sec:introduction}
Majorana bound states (MBS) have been of great interest for quantum computing over the past two decades due to their non-Abelian statistics and robustness against local perturbations\cite{nayak:rmp08, alicea:rpp12}. Different models for creation of MBS have been suggested and studied\cite{nayak:rmp08, alicea:rpp12, kitaev:pu01, fu:prl08, tanaka:prl09, wilczek:natphys09, akhmerov:prl09, franz:physics10, stern:nature10, sau:prl10, lutchyn:prl10, oreg:prl10, alicea:prb10, flensberg:prb10, duckheim:prb11, klinovaja:prl12, kjaergaard:prb12, turcotte:arxiv19, mourik:science12, deng:nanolett12, das:natphys12, rokhinson:natphys12, finck:prl13, churchill:prb13, nadjperge:science14, zhang:natcom17, deng:science16, zhang:nature18, higginbotham:natphys15, albrecht:nature16}. One of the models, which has attracted much attention because of its potential experimental feasibility, is a nanowire-superconductor hybrid system\cite{lutchyn:prl10, oreg:prl10}. It is constructed from a nanowire with strong spin-orbit interactions in a uniform magnetic field
%, acting on different, orthogonal spin components, and 
on a superconducting substrate, which induces superconductivity in the nanowire due to the proximity effect. The Hamiltonian for the semiconducting part of this device, i.e.\ the nanowire with spin orbit interaction and a uniform magnetic field, is related by a unitary transformation to a Hamiltonian for a nanowire with a helical magnetic field and no spin orbit interaction~\cite{braunecker:prb10}. The presence of MBS in nanowires and carbon nanotubes with helical magnetic fields was studied in Refs.~\onlinecite{klinovaja:prl12, egger:prb12}. It was also suggested to create a similar setup with a helical-shaped effective magnetic field via magnetic atoms on top of a superconductor\cite{choy:prb11, vazifeh:prl13, braunecker:prl13, nadjperge:prb13, nadjperge:science14, pientka:prb13, ruby:prl15}. Non-uniform magnetic fields, created by an array of nanomagnets, can be used to create MBS in a nanowire-superconductor hybrid system~\cite{kjaergaard:prb12,turcotte:arxiv19}. The formation and braiding of MBS via a nanomagnet pattern on a 2D substrate was discussed in Refs.~\onlinecite{fatin:prl16, matos:ssc17}. There are other suggestions for devices with various magnetic field shapes and origins which may host MBS, e.g., Refs.~\onlinecite{ojanen:prb13, sedlmayr:prb15, boutin:arxiv18}.

Recent work in Ref.~\onlinecite{maurer:arxiv18} presented detailed modelling of the magnetic field due to an array of nanomagnets acting on a nanowire in a Si heterostructure. As Si is widely used in modern technology, and therefore a material convenient for potential applications, it can be useful for an experimental realization of MBS to study whether certain Si structures can host MBS. Here, we consider a Si nanowire with superconductivity induced by the proximity effect and with nearby nanomagnets that can be made out of Co or SmCo\cite{maurer:arxiv18}. %The Si nanowire can be made via gating or etching of a two-dimensional electron gas in a Si-based heterostructure. 

\begin{figure}[tb]
\centering
\includegraphics[width=0.95\linewidth]{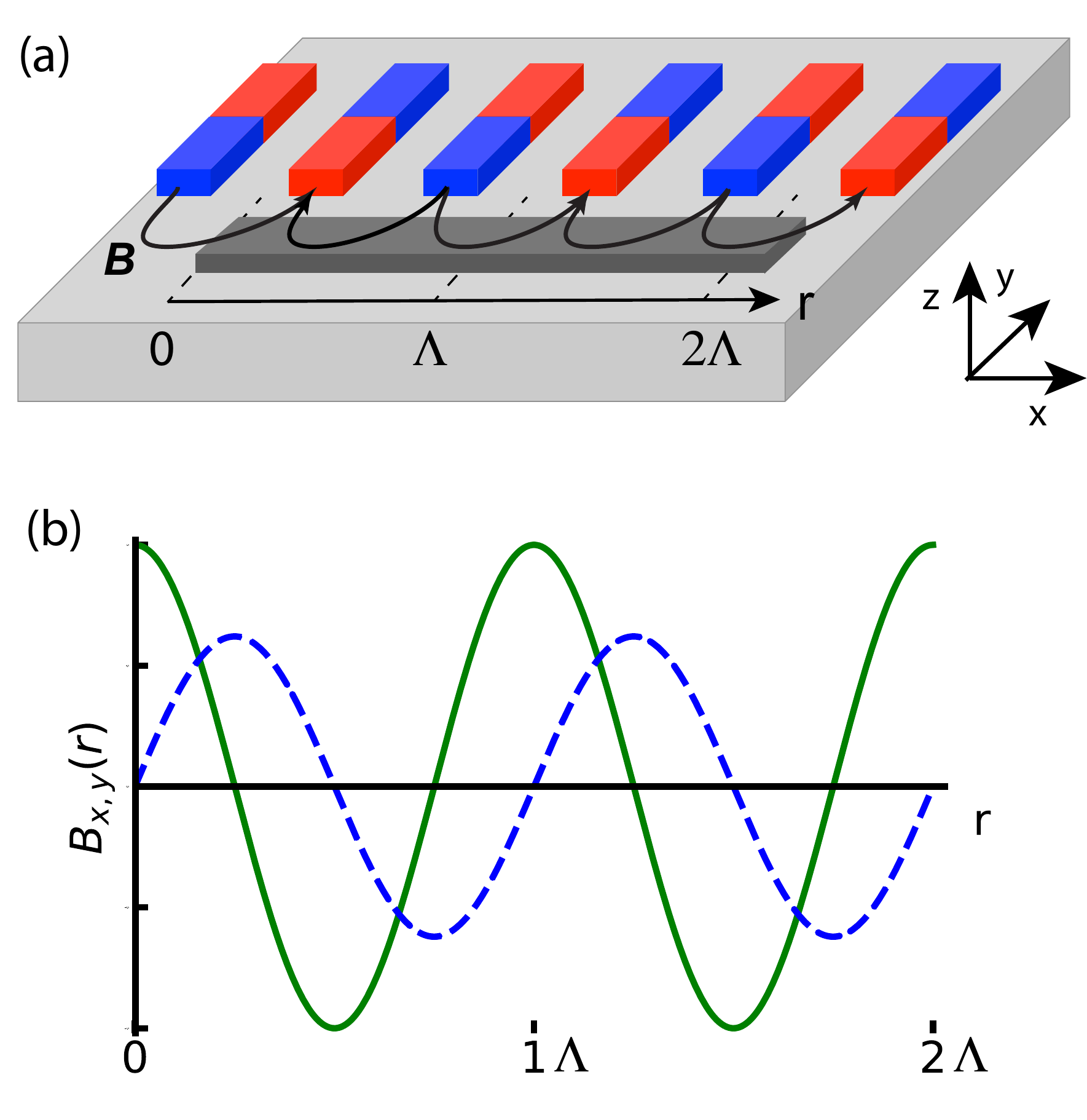}
\caption{(a) A schematic representation of the system geometry. Here, the nanowire (orange cylinder) is in proximity with a superconductor (grey rectangle). Nanomagnets with alternating magnetization are arranged nearby (blue-red rectangles), in the same plane as the wire. 
%Note that the magnetization of the nanomagnets does not have to be alternating, but might have other configurations with different positioning of the magnets. 
Alternatively, the nanomagnets can be positioned higher than the nanowire, which could improve the shape of the magnetic field for our purposes\cite{maurer:arxiv18}. (b) Components of the spiral magnetic field, $B_x=B_{x0}\cos(2\pi r/\Lambda)$ (solid line) and $B_y=B_{y0}\sin(2\pi r/\Lambda)$ (dashed line) in the nanowire, with period $\Lambda$, as a function of position along the wire, $r$. For an ideal helical field, we have $B_{y0}=B_{x0}$. Here, we show the case of an elliptic helical field, with $B_{y0}=0.62B_{x0}$.
% (c) FIX CURVES TO CORRESPOND THE ACTUAL CHOICE Magnetic field components $B_x$, $B_y$, $B_z$ as a function of the position along the wire $r$ for a spirally shaped magnetic field with elliptical cross-section. (d) REMOVE: Magnetic field components $B_x$, $B_y$, $B_z$ as a function of $r$ for a spatially varying magnetic field similar to the one presented in Ref.~\onlinecite{maurer:arxiv18}. 
}
\label{fig:setup_Majoranas}
\end{figure}

In this work, we investigate when a topological superconducting phase in lithographically defined Si nanowires exists.  
Using parameters that are reasonable for lithographically defined silicon nanowires and magnets (see Sec.~\ref{subsec:experimental_parameters}), we consider
25-nm wide wires with a superconducting gap $\sim 5~\mu$eV and with the magnetic field produced by nanomagnets with strength about $100$~mT, see Fig.~\ref{fig:setup_Majoranas}.  These conditions are sufficient for a perfect helical magnetic field to produce the topologically non-trivial superconducting phase that supports an MBS with localization length of about $1~\mu$m%, this result is in agreement with previous theory of the MBS in helical magnetic field
~\cite{braunecker:prb10,klinovaja:prl12, egger:prb12}.
Since an ideal helix is difficult to achieve using micromagnets,  we study how different shapes of the magnetic field would affect the presence of an MBS in a Si-based setup. In particular, we consider a spiral magnetic field with an elliptic cross section. For both ideal and non-ideal helical fields, a partial gap opens in the presence of a magnetic field.
However in the non-ideal case, a second gap opens that is proportional to the difference between the major and minor axes of the spiral~\cite{maurer:arxiv18}.
If the chemical potential is tuned such that it is inside both gaps, the
superconductivity and the MBS are both suppressed. 
%This can, however, be modified by tuning the chemical potential closer to the smaller magnetic gap edge. 

We investigate the phase boundary of the topological superconducting phase as a function of the major and minor axes of the spiral elliptic magnetic field.  The boundary between topological and non-topological phases is marked by 
the vanishing of the superconducting gap around the chemical potential.  
In the non-topological phase, there are no states below the gap, while in the topological phase two additional states, the MBS localized at the wire edges, develop.  The localization length of the MBS decreases quickly as the superconducting gap recovers away from the phase boundary.  Thus, we use the existence of two states with eigenenergies below the superconducting gap, together with their localization near edges, as a criterion for the topological phase.  We demonstrate that the eigenenergies of the topological state are exponentially suppressed for long enough wires, yielding effective zero-energy modes that are associated with the topological superconducting phase.

MBS develop in the presence of a perfect helical magnetic field when the field magnitude exceeds a threshold value equal to the superconducting order parameter in the wire~\cite{braunecker:prb10,klinovaja:prl12, egger:prb12}.  A small deformation of the perfect helix is not expected to immediately destroy the MBS.  Our analysis demonstrates the robustness of the MBS to relatively strong deformation of the helical field.
%, as the phase diagram shows the presence of MBS in the system for a wide range of ellipticity of the magnetic field.  
As we demonstrate below, when one amplitude of the oscillating magnetic field is  tuned so that it is about twice the value of the superconducting order parameter in the wire, the MBS develop even when the second component of the field is much smaller;
% weaker second component of the field, 
thus, when the magnitude of one of the field components is tuned appropriately, the MBS can survive even very strong ellipticity of the helical magnetic field.
We also show that the analytic solution of a continuum model with a linearized energy dispersion provides a good guide for understanding the
numerical results obtained for finite, discretized wires.

The paper is organized as follows. In Sec.~\ref{sec:model} %\textbf{needs an update} 
we discuss the experimental constraints on the model parameters for an example system of a nanowire in which the superconductivity is induced by the proximity effect and magnets are patterned lithographically.  We then present the Hamiltonian that we analyze in the succeeding sections, \ref{subsec:superlattice} and~\ref{subsec:superconductivity_in_Hamiltonian}. We present an analytical derivation of the MBS wave function and the spectrum for a spiral magnetic field with an elliptic cross-section in Sec.~\ref{sec:AnalyticalConsideration}.   Sec.~\ref{sec:DiscretizedHamiltonian} presents numerical results for the phase diagrams for the different shapes of the magnetic field. %In Sec.~\ref{sec:HigherHarmonics} we discuss the effect of higher harmonics of the magnetic field on the MBS wave function. 
Our conclusions follow in Sec.~\ref{sec:conclusions}. 
%Details of the calculations and additional information are provided in the Appendices.

%
\section{Model of superconducting nanowire in periodic magnetic field}
\label{sec:model}

\subsection{Estimations of experimental parameters}
\label{subsec:experimental_parameters}
%\red{Although the focus of this work is theoretical, it is important to note that the physical requirements are realistic. 
%To highlight this point, we consider a specific model system, for which the experimental parameters are estimated below.
%However, we wish to emphasize that our analysis is not limited to this specific physical implementation.}

%To build a nanowire-superconductor system with nanomagnets, we consider the following representative structure: we start with a Si \st{substrate with high Ga or B doping to induce superconductivity}\cite{heera:njp13, bustarret:nature06}.
%\red{arranged in direct proximity with a bulk superconductor. 
%The wire may be fabricated, preferably by lithographic deposition of appropriate top gates, or, if necessary, by etching.}
%\st{Atop this layer we fabricate a Si wire (preferably by lithographic deposition of appropriate top gates, or, if necessary, by etching). If the superconductivity is proximity-induced, }
%It is important to avoid a Schottky barrier between the wire and the superconductor~\cite{klogstrup:2015}. Lithographically defined Co and SmCo nanomagnets\cite{maurer:arxiv18} deposited nearby give rise to appropriate helical field variations, as shown in Fig.~\ref{fig:setup_Majoranas}. 
%Below, we provide estimates of several experimental parameters that are used in our calculations.

While the focus of this work is theoretical, it is important to note that the physical regimes are realistic.  An example physical system is a silicon nanowire, whose width we estimate below, with superconductivity induced by the proximity effect, either from metals~\cite{klogstrup:2015} or from the superconductivity in a nearby, very highly doped semiconductor region~\cite{bustarret:nature06}.  Lithographically defined Co and SmCo nanomagnets\cite{maurer:arxiv18} deposited nearby give rise to appropriate helical field variations, as shown in Fig.~\ref{fig:setup_Majoranas}. 

We now estimate the transverse width of the nanowire and its associated Fermi wavelength.
Because the experimental parameters of interest (e.g., the threshold density) are better characterized in two-dimensional (2D) systems than in wires, we will refer to 2D experiments as a starting point.
The six-fold valley degeneracy of the conduction band in bulk silicon is lifted by tensile strain or by narrow confinement, leaving just two low-energy valleys to form a quantum device\cite{Friesen2007}.
The remaining degeneracy is lifted by wavefunction overlap with sharp interfaces, with a valley energy splitting of $\delta_v$.
Assuming a parabolic dispersion relation for the two-dimensional electron gas (2DEG), the lower ($l$) and upper ($u$) valley band energies are given by
 $\varepsilon_l(\mathbf{p}) = p^2/2m$ and $\varepsilon_u(\mathbf{p}) = p^2/2m+\delta_v$, where  $\mathbf{p}$ is a two-dimensional quasi-momentum and $m=1.73\times 10^{-31}$ kg is the transverse effective electron mass.
Here, we choose $\delta_v\lesssim 100\mu$eV as the valley splitting of a typical 2DEG\cite{gamble:apl2016}, although there is some evidence of larger valley splittings in wire geometries, depending on the confinement\cite{Ibberson2018}.

The Fermi energy $E_F$ should be large enough to allow the nanowire to conduct, where $E_F$ is measured from the bottom of the lower valley in the 2DEG dispersion. Normally the threshold electron density needed for a 2DEG to conduct is smaller than for a wire, since any disorder disrupts the current flow in the one-dimensional case. For a nanowire, we therefore assume an electron density $n_e$ for which the Fermi energy is higher than the maximum value of the disorder potential. The Fermi energy and the electron density are then related by  
\begin{equation}
n_e = \frac{2 m (2E_F-\delta_v)}{2\pi\hbar^2},
\end{equation}
assuming spin degeneracy.

To determine the size of the nanowire, we assume a harmonic confinement potential in the transverse direction with a root-mean-square width of the wavefunction, $\sigma_w$, corresponding to an energy level splitting of 
\begin{equation}
\hbar \omega_0  = \frac{\hbar^2}{m\sigma_w^2}.
\end{equation}
Since the valley degree of freedom represents an unwanted quantum variable, we can suppress the filling of the upper valley band by adjusting the ground-state energy $\hbar\omega_0/2$ such that it lies between the highest filled state and the lowest unfilled state:
\begin{equation}
E_F - \delta_v< \frac{\hbar\omega_0}{2} < E_F.
\end{equation}
To satisfy this constraint, we adopt $\sigma_w=25$~nm, yielding an upper limit of $n_e=5.9\times 10^{10}$cm$^{-2}$ for the electron density, which as desired is significantly higher than the threshold electron density of a conducting 2DEG, $n_{e,\text{th}}=2\times 10^{10}$ cm$^{-2}$, as reported in Ref.~\onlinecite{laroche:aipadv15} for a 100 nm deep Si/SiGe quantum well.

The chemical potential in the wire is counted from the bottom of the one-dimensional conduction channel, such that
$$
\mu = E_F - \frac{\hbar^2}{2m\sigma_w^2},
$$ 
with a corresponding Fermi wavevector of $k_F = \sqrt{2m\mu/\hbar^2 }$.  
Here we choose 
$\mu\simeq 50$~$\mu$eV, so that the occupation of the higher valley is well suppressed.  

Finally, we estimate the values of the magnetic field $B$ and the proximity-induced superconducting gap in the nanowire $\Delta$ that support an MBS.  For a perfectly helical magnetic field, the presence of an MBS requires fields with $g\mu_BB > \sqrt{\Delta^2 + \delta\mu^2}$\cite{klinovaja:prl12,kjaergaard:prb12}, where the Land\'{e} $g$-factor $g\approx 2$ for Si, $\mu_B$ is the Bohr magneton,  and $\delta \mu = \mu - \hbar^2Q^2/2m$ is the detuning of the chemical potential away from the center of the energy gap ($\hbar^2Q^2/2m$), caused by a magnetic superlattice with period $\Lambda$ (see Fig.~\ref{fig:setup_Majoranas}) and wavevector $Q=2\pi/\Lambda$.
(Henceforth, we adopt energy units for $B$ by absorbing $g\mu_B$ into its definition.)
% \snccomment{I deleted a mention of the appendix (I am voting for deleting both appendices).}  
 Because it is difficult to achieve Zeeman splittings in excess of 20~$\mu$eV using nanomagnets,
we take $\Delta=5~\mu$eV here. This choice also satisfies the condition $\mu \gg \Delta$, which is necessary for achieving a proximitized superconducting gap in the wire.

%We also take $g_s=g_v=2$ and obtain
%\begin{eqnarray}
%E_F&=&0.25\  {\rm meV},\\
%\lambda_F&=&177\  {\rm nm},\\
%\sigma_0&=&28\  {\rm nm}.
%\end{eqnarray}

\begin{figure}[tb]
\begin{center}
\includegraphics[width=1\linewidth]{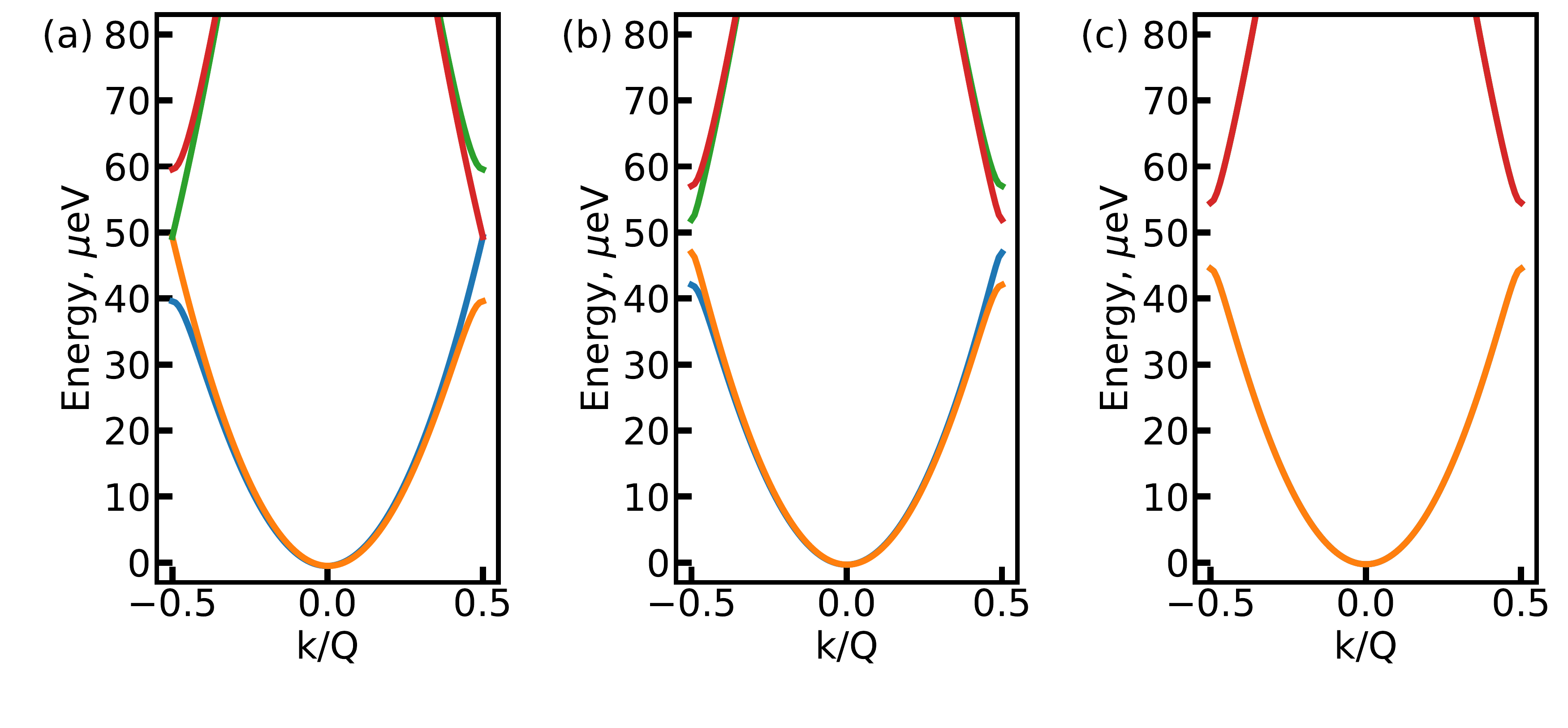}
\caption{ The band structure (energy versus normalized wavevector $k/Q$) of an electron in the presence of the periodic, helical magnetic field, defined in Eq.~\eqref{eq:BxBy}, with period $\Lambda=2\pi/Q=200$~nm. 
(a) An ideal helical magnetic field $B_{x0}=B_{y0}=10~\mu$eV.
(b) An elliptical magnetic field $B_{x0}=10~\mu$eV and $B_{y0}=5~\mu$eV. 
(c) A non-chiral magnetic field $B_{x0}=10~\mu$eV and $B_{y0}=0$.
\snccomment{Need to say what the colors are.  Here is my guess:}  
States with the two spin chiralities with energies below the gap  are denoted in orange and blue, while states with the two spin chiralities with energies above the gap are denoted in red and green.
\snccomment{Rest of caption:}
For the ideal helical field (a), a gap opens for one spin chirality but not the other, so there are states in at least one band for all values of the chemical potential.  When the field is non-chiral (c), the energy dispersions of the different spin helicities are the same, and there is no chemical potential for which one band is gapped and the other is not.  (The blue and green curves are not visible on the plot because they are identical to the orange and red curves.)    When the field is chiral but not an ideal helix (b), the gaps of the two helicities are different, and there are values of the chemical potential that are in the gap for one chirality but not the other.
}
\label{fig:magnetic_bands}
\end{center}
\end{figure}

\subsection{Magnetic superlattice}
\label{subsec:superlattice}

In the following two sections, we present the Hamiltonian studied in this work.  First, we introduce a periodic magnetic field in the absence of superconductvity. In Sec.~ \ref{subsec:superconductivity_in_Hamiltonian}, we then include the effects of superconductivity.
%$\sigma$ and $\tau$ are Pauli matrices in spin and particle-hole spaces, respectively. The coordinate along the nanowire is $r$, and the effective mass of an electron is $m$. 
%We propose using \snc{a Si-based structure consisting of} a Si nanowire and a \snc{doped} substrate that induces superconductivity, combined with Co and SmCo nanomagnets, see Fig.~\ref{fig:setup_Majoranas}.  

We consider
the Hamiltonian for a single electron in the wire with parabolic energy dispersion in the presence of a magnetic field $\mathbf{B}(r) = \{B_x(r), B_y(r), B_z(r)\}$  that oscillates periodically as a function of the coordinate $r$ along the wire:
\begin{eqnarray}
\label{eq:HZ}
H_Z=\varepsilon_l(\mathbf{p})+\mathbf{B}(r)\cdot{\bm{\sigma}} -\mu
%\frac{1}{2} [(B_{x0}-B_{y0})(R^\dagger_\uparrow L_\downarrow+L_\downarrow^\dagger R_\uparrow)+\\ \nonumber+(B_{x0}+B_{y0})(L_\uparrow^\dagger R_\downarrow+R_\downarrow^\dagger L_\uparrow)]
,
\end{eqnarray}
where $\bm{\sigma} = \{\sigma_x,\sigma_y,\sigma_z \}$ are Pauli matrices.

The actual magnetic field configuration produced by nanomagnets is complex.  Field configurations produced by arrays of bar-shaped nanomagnets as well as electron spectra are calculated in Ref.~\onlinecite{maurer:arxiv18}. It was also shown that a special configuration of magnetic fields may improve conditions for the MBS to develop.  However, the goal of this paper is to investigate how deviations from the perfect helical magnetic field may affect the topologically non-trivial superconducting phase.  For this purpose, we 
use a helical magnetic field with elliptical helical cross-section:
\begin{equation}
B_x=B_{x0}\cos{Q r},\quad B_y=B_{y0}\sin{Q r},\quad B_z=0,
\label{eq:BxBy}
\end{equation}
where $Q=2\pi/\Lambda$ is the vector of the reciprocal 1D Brave lattice of the magnetic superlattice with period $\Lambda$ and $B_{x0},B_{z0}\geq 0$.
We note, that due to the absence of the spin--orbit interaction in this system, the direction of the wire and magnetic field orientation are completely decoupled.
For example, the system properties remain the same regardless of the choice of the magnetic components $x,y$ relative to the wire direction.

For the magnetic field given by Eq.~\ref{eq:BxBy}, the matrix elements of the magnetic periodic potential are non-zero only for two reciprocal vectors  $\pm Q$:
\begin{equation}
W_{\pm Q} =\frac{1}{2} \left(
\begin{array}{cc}
0 & B_{x0}\pm B_{y0} \\ 
B_{x0}\mp B_{y0} & 0
\end{array} 
\right).
\end{equation}
The spectral equation for electron states in the magnetic superlattice has the form
\begin{equation}
\left(
\frac{\hbar^ 2(k-nQ)^2}{2m} -{\cal E}(k)\right) \bm{c}_{k-nQ} +\sum_{\pm} W_{\pm Q}
\bm{c}_{k-(n\pm 1)Q}=0,
\label{eq:magnetic_bands}
\end{equation}
where $n =0,\pm 1, \pm 2\dots$ and the spinor coefficients $\bm{c} _{k-nQ}$ define the electron wave function $\bm{\psi}_k(r) = \sum_{n} \bm{c}_{k-nQ}\exp(i(k-nQ)r)$~\cite{maurer:arxiv18,ashcroft:1976}.
%\snccomment{It would probably be better to use a letter other than U for the potential because U is used for a unitary matrix later on.  Can we use V instead?}

Using Eq.~(\ref{eq:magnetic_bands}), we can determine that there are energy gaps at the edges of the Brillouin zone with magnitudes %gaps are given by 
$\left |B_{x0}\pm B_{y0} \right |$.
%There is always a large gap $B_{x0}+B_{y0}$ for one branch of the spectrum.
For an ideal helical field with $B_{x0}=B_{y0}$, one branch has a large gap $B_{x0}+B_{y0}$, while the other branch is gapless. However, when $B_{x0}\neq B_{y0}$, both gaps are nonzero, and there are no states within the energy window $|B_{x0}-B_{y0}|$ around $\epsilon = \hbar^2Q^2/8m$.
%\snc{When the superconducting pairing energy $\Delta$ is small and the chemical potential $\mu$ is within this complete gap, superconductivity is suppressed, which in turn suppresses the topologically non-trivial superconducting phase.  However, as we show below, when $\Delta$ is comparable to this gap, then robust topologically nontrivial superconductivity can be supported even when the chemical potential $\mu$ is in this gap}. 

We solve Eq.~(\ref{eq:magnetic_bands}) for for the magnetic field period $\Lambda = 200$~nm in Si nanowire.
The energy bands are shown in Fig.~\ref{fig:magnetic_bands}(a) for 
$B_{x0}=B_{y0}=10~\mu$eV, in Fig.~\ref{fig:magnetic_bands}(b) for 
$B_{x0}=2B_{y0}=10~\mu$eV, and in Fig.~\ref{fig:magnetic_bands}(c) for
$B_{x0}=10~\mu$eV, $B_{y0}=0$.
For the last case, the magnetic field is non-chiral, the two spin helicities have identical band structures, and
no topologically nontrivial phase is supported.

Using parameters from Subsec.~\ref{subsec:experimental_parameters}, 
we find that $\Lambda = 200$~nm is an acceptable scale for nanofabrication and at the same time allows the lifting of the valley degeneracy of the conducting channel in the wire.  Indeed, the magnetic structure would require fabrication of pairs of 50~nm wide nanomagnets with opposite magnetizations, see Fig.~\ref{fig:setup_Majoranas}.

%\textbf{Keep, unless we remove this section}In Subsec.~\ref{subsec:HigherHarmonics} we also consider the presence of the higher order harmonics of the magnetic field with the wave numbers $2q\kappa$, where $q>1$, $q\in \mathcal{N}$ in order to take into account possible imperfections of the sinusoidal form. \snc{I think it is reasonable to include a short discussion of how the calculation changes when the harmonics is included (specifically, write down the analog of (6) and (7) when the harmonic is included), then mention that additional gaps open, but since none of them are in the relevant energy range they are not expected to  affect things much, and that this conclusion is consistent with the numerical results shown later.}
%\snc{I think we can delete this sentence: For convenience, we take $B_{x0},B_{y0},B_{z0},\Delta\geq 0$ in this work. }

\subsection{Hamiltonian for superconducting wire}
\label{subsec:superconductivity_in_Hamiltonian}

We now add to the Hamiltonian the superconductivity terms 
%to the Hamiltonian of the system.  The proximity effect introduces pairing of 
that pair electrons with energies above the chemical potential with holes below the chemical potential.  This coupling is conveniently represented by the electron creation and annihilation operators in the Nambu space defined by the vector
$\bm{\Psi}=\{\psi_\uparrow, \psi_\downarrow, \psi_\uparrow^\dagger, \psi_\downarrow^\dagger\}^T$.
The Hamiltonian of the system can be written as 
\begin{subequations}
\begin{eqnarray}
\label{eq:MainHamiltonian_integral}
\mathcal{H}&=&\int  \bm{\Psi}^\dagger(r) H\bm{\Psi}(r) dr,
\end{eqnarray}
where the Hamiltonian matrix in the Nambu space is
\begin{eqnarray}
\label{eq:MainHamiltonian_integral2}
H&=&\left(\begin{array}{cc}
H_Z & i\Delta\sigma_y \\ 
-i\Delta\sigma_y & -H_Z^*
\end{array} 
\right)~,
\end{eqnarray}
where $r$ is the position along the wire, $\Delta$ is the superconducting gap,
 and the single-electron Hamiltonian $H_Z$ is given by Eq.\ \eqref{eq:HZ}.
\end{subequations}
This Hamiltonian has eigenvectors that are solutions to the Bogolyubov-De Gennes (BdG) equation~\cite{deGennes:1966}
\begin{equation}
\label{eq:BdG}
H \bm{\Psi} = E\bm{\Psi}.
\end{equation}
Below we will investigate the eigenenergies and eigenstates of this Hamiltonian by finding solutions of the BdG equation numerically.

The electron wavefunctions can be rewritten in the Majorana basis instead of the Nambu basis by introducing a 
unitary transformation of the vector $\Psi$, where the  transformation matrix is given by
\begin{equation}
\label{eq:MajoranaBasis}
\bm{\Phi} =  U_M\bm{\Psi}, \quad 
U_M=\frac{1}{\sqrt{2}}\begin{pmatrix}1&0&1&0\\ 0 & 1 & 0 & 1\\ i&0&-i&0\\ 0 & i& 0 &-i\end{pmatrix}~.
\end{equation}
The MBS in this basis is represented by a real function with eigenenergy  $E_0\to 0$.  

%\snccomment{I moved the paragraph discussing the size of magnetic field and superconducting gap to II.A.}

\section{Analytical Characterization of the Phase Diagram}
\label{sec:AnalyticalConsideration}
%\subsection{Linearized energy spectrum}
To provide analytic insight into when a magnetic field that does not have an ideal helical form induces MBS, we 
%study analytically a spiral magnetic field with an elliptical cross section. We 
generalize the procedure described in Ref.~\onlinecite{klinovaja:prb12} to apply to the case of a field with an elliptical cross section.  We consider perfect matching between the Fermi momentum of 1D  electrons in the wire and the periodicity of the magnetic superlattice by setting the chemical potential $\mu = \hbar^2 Q^2/8m$.  
We choose $B_{x0}>0$ and $B_{y0}>0$; this restriction is inessential because different signs of these components correspond to different chiralities of the magnetic field.
%\snccomment{I don't understand what this next sentence is trying to say.}
%The energy spectrum of the system is similar to the spectrum for the ideal helical field described in Ref.~\onlinecite{braunecker:prb10}\snc{;}
% using a continuum model
% (\snc{??}see also Appendix \ref{app:SpectrumHelicalField}), 
 %which should be \snc{valid} if the period of magnetic field oscillations, $2\pi/Q$, is small enough, i.e., in the limit $B_{x0},B_{y0},%\Delta\ll \mu$. 
%\snccomment{My guess is: 
The continuum and long-length limits examined here are expected to be applicable when the period of the magnetic field oscillations is
much less than the length of the wire and when $B_{x0},B_{y0},\Delta\ll \mu$.
%}

We represent the electron wave functions as a superposition of  left- and right-movers
\begin{equation}
\psi_\sigma=R_\sigma e^{iQ r/2}+L_\sigma e^{-iQ r/2},
\end{equation}
where $\sigma=\{\uparrow,\downarrow\}$, and we linearize the energy dispersion in the kinetic energy term:
\begin{equation}
H_{\rm kin}=-i\hbar v_F (R_\uparrow^\dagger\partial_r R_\uparrow-L_\uparrow^\dagger\partial_r L_\uparrow+R_\downarrow^\dagger\partial_r R_\downarrow-L_\downarrow^\dagger\partial_r L_\downarrow),
\end{equation}
where the Fermi velocity is $v_F = \hbar Q/2m$.
The hole part of the Hamiltonian can be obtained from the anticommutation relations of the fermion operators and we do not explicitly show it here. We neglect all fast-oscillating terms, assuming that the localization length of the $R$ and $L$ functions is much larger than $2\pi/Q$. Later we show that this condition indeed holds for our results. 
The Zeeman term due to the magnetic field is
\begin{eqnarray}
H_{\rm mag}=\frac{1}{2} [(B_{x0}-B_{y0})(R^\dagger_\uparrow L_\downarrow+L_\downarrow^\dagger R_\uparrow)+\\ \nonumber+(B_{x0}+B_{y0})(L_\uparrow^\dagger R_\downarrow+R_\downarrow^\dagger L_\uparrow)],
\end{eqnarray}
and the superconductivity term 
\begin{equation}
H_{sc}=\Delta (R_\downarrow L_\uparrow+L_\downarrow R_\uparrow-R_\uparrow L_\downarrow-L_\uparrow R_\downarrow+H.c.).
\end{equation}
The Hamiltonian can be decoupled into two non-interacting subspaces $\bm{\chi}_-=\{R_\uparrow, L_\downarrow, R_\uparrow^\dagger, L_\downarrow^\dagger\}^T$ and $\bm{\chi}_{+}=\{L_\uparrow, R_\downarrow, L_\uparrow^\dagger, R_\downarrow^\dagger\}^T$. In these subspaces, the Hamiltonian has the form
%\begin{subequations}
\begin{eqnarray}
%\label{eq:H1}
\label{eq:Hchiral}
H_\pm&=&\pm i\hbar v_F \partial_r \sigma_z -\delta\mu  \tau_z +B_\pm\sigma_x\tau_z-\Delta\sigma_y\tau_y,
%\\
%\label{eq:H2}
%H_\circlearrowleft&=&i\hbar v_F\partial_r \sigma_z+B_\circlearrowleft \sigma_x\tau_z-\Delta \sigma_y\tau_y,
\end{eqnarray}
%\end{subequations}
where $B_\pm =(B_{x0}\pm B_{y0})/2$ and $\delta \mu =\mu-\hbar^2Q^2/8m$ is the mismatch between the chemical potential and the center of energy gap of electron bands of the magnetic superlattice.
% and $B_\circlearrowleft = (B_{x0}+B_{y0})/2$.
The energies of quasiparticle excitations of $H_\theta$ with $\theta=\pm$ for perfect matching of the chemical potential, $\delta\mu=0$, are given by
%\begin{subequations}
\label{eq:linspectrum}
\begin{eqnarray}
\label{eq:spectrum}
E_\theta^\pm=\sqrt{\hbar^2 v_F^2 \delta k^2+\left[B_\theta\pm\Delta\right]^2},
\label{eq:spectrum2}
%E_{II}^\pm=\sqrt{v_F^2 \delta k^2+\left[\frac{1}{2}(B_{x0}+B_{y0})\pm \Delta\right]^2},
\end{eqnarray}
%\end{subequations}
where $\delta k = k \pm Q/2$ denotes momenta counted from Fermi points $\pm Q/2$ and we consider only non-negative energies. 
%Here we consider the case of $B_{x0}\geq B_{y0}$; the analysis for the opposite case is similar and the result will be presented explicitly below.
%in the opposite case we can relabel axes that also flips the sign of the first term in Eq.~\eqref{eq:Hchiral}.
We note that the two subspaces, $\theta =\pm$, describe two chiralities of electrons in the spiral magnetic field.  Without the proximity effect, $\Delta=0$, Eqs.~\eqref{eq:linspectrum} correspond to two branches with a smaller and larger magnetic gaps at the boundary of the Brillouin zone of the magnetic superlattice, see Subsec.~\ref{subsec:superlattice} and Fig.~\ref{fig:magnetic_bands}.

\begin{figure}[tb]
\begin{center}
\includegraphics[width=\linewidth]{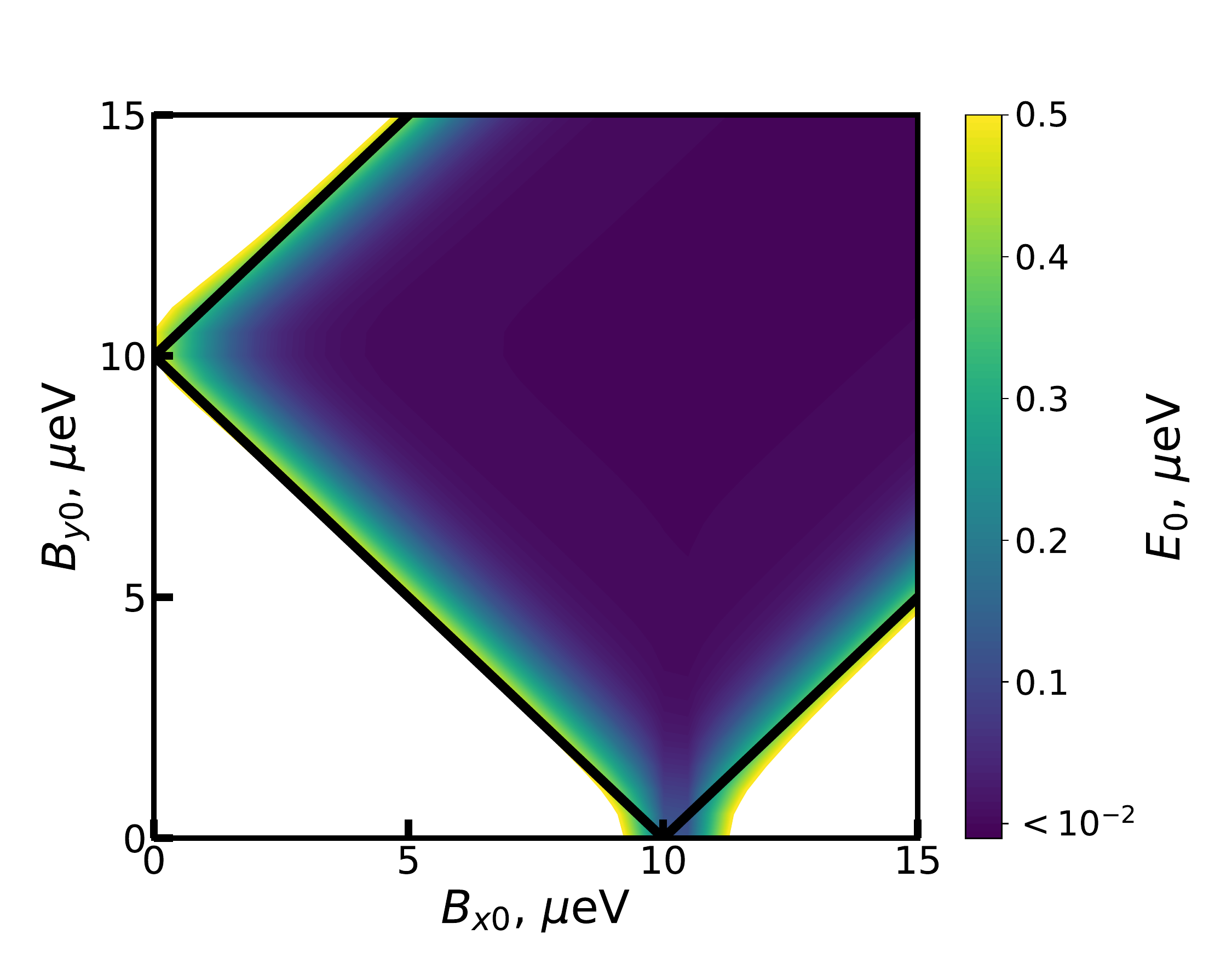}
\caption{Contour plot of $E_0$, the energy ofthe state of lowest positive energy in the system, versus the magnetic field amplitudes $B_{x0}$ and $B_{y0}$. Parameter values used are: superconducting order parameter $\Delta = 5~\mu$eV,  magnetic field wavevector $Q = 2 \pi/\Lambda$,  $\Lambda = 200$~nm, chemical potential $\mu=\hbar^2Q^2/8m = 49.6~\mu$eV,  and wire length $L=20~\mu$.  
%\snc{The numerical results are obtained using a discretization length $\delta r=20~$nm.} 
The uncolored region where $E_0 > 0.5~\mu$eV corresponds to the gapped non-topological superconducting phase,  
the dark purple color denotes the region in which $E_0 < 10^{-2}\mu$eV, and
the crossover region over which the zero energy state develops in this wire with finite length is evident in the transitions between colors in the color plot for the region of $E_0<0.5\mu$eV.
The solid straight lines show the analytical predictions for the phase boundary given by Eq.~\eqref{eq:transitions_zeromu}.
}
\label{fig:phase_diagram_mu}
\end{center}
\end{figure}

We observe that the excitation energy vanishes for $\delta k=0$ when 
\begin{equation}
\label{eq:phase_boundary_analytical}
{\mathrm {either}~} B_+ = \Delta~{\mathrm {or}}~ B_- = \Delta~.
\end{equation}  
These lines, which define the boundaries between topologically trivial and non-trivial superconducting phases in an infinitely long wire, are shown as solid straight lines in Fig.~\ref{fig:phase_diagram_mu}.  Our next step is to demonstrate that the internal region in the phase diagram indeed supports the MBS.

Away from the lines defined by Eq.~\eqref{eq:phase_boundary_analytical}, there is no zero-energy eigenstate for real $\delta k$.  However, the purely imaginary values of 
\begin{equation}
\label{eq:evanescent_wave}
\delta k^{\pm}_{\alpha,\theta} = \pm i\frac{\alpha B_\theta + \Delta}{\hbar v_F}
\end{equation} 
might describe zero-energy solutions that exponentially decrease or grow along the nanowire.  The $\pm$ sign in Eq.~\eqref{eq:evanescent_wave} defines the solutions that decrease or increase as a function of coordinate $r $ along the wire and would correspond to two states localized at each of the ends of the nanowire, and $\alpha = \pm$ identifies the sign choice in Eq.~\eqref{eq:spectrum}.  Here we focus on the solution that is localized near $r=0$.  In this case, we identify only one pair of $\delta^{\pm}_{\alpha,\theta=+}$ and $\delta^{\pm}_{\alpha,\theta=-}$ that satisfy the boundary condition $\bm{\Psi}(r=0)=0$.

\mgvcomment{Equations and explanation of the next three equations were corrected and the explanation is modified:}
The general solution can  be written as a linear combination of eight terms 
\begin{equation}
%\chi(r) = \alpha_+ \chi_+(r) + \alpha_- \chi_-(r), 
%\end{equation}
%where 
%\begin{equation}
\label{eq:chi}
\bm{\chi}(r) = \sum_{\theta=\pm} \sum_{\alpha=\pm}\sum_{\sigma=\pm}
\beta^{\sigma}_{\alpha,\theta} \bm{c}^{\sigma}_{\alpha,\theta} 
\exp(ir\delta k^{\sigma}_{\alpha,\theta})
\end{equation}
of four linearly independent  $4-$component vectors %$c^{\sigma}_{\alpha,\theta}$ are 
\begin{equation}
\begin{split}
&\bm{c}^{+}_{-,-} = \bm{c}^{-}_{-,+} = \left(\begin{array}{c}
1 \\ 
i \\ 
1 \\ 
-i
\end{array} 
\right),\quad
\bm{c}^{+}_{+,-} =\bm{c}^{-}_{+,+} = \left(\begin{array}{c}
1 \\ 
-i \\ 
-1 \\ 
-i
\end{array} 
\right),\\
&
\bm{c}^{-}_{-,-} =\bm{c}^{+}_{-,+} = \left(\begin{array}{c}
1 \\ 
-i \\ 
1 \\ 
i
\end{array} 
\right),\quad
\bm{c}^{-}_{+,-} =\bm{c}^{+}_{+,+} =  \left(\begin{array}{c}
1 \\ 
i \\ 
-1 \\ 
i
\end{array} 
\right).
\end{split}
\label{eq:4vecs}
\end{equation}

The proper solution \eqref{eq:chi} vanishes at $r=0$, so it must contain a pair of terms formed by one of the vectors of Eq.~\eqref{eq:4vecs} and multiplied by the exponential functions $\exp(ir\delta k^{\sigma}_{\alpha,\theta})$ with different top index $\sigma$ in $\delta k^{\sigma}_{\alpha,\theta}$. 
At the same time, each term in the pair must decrease as a function of $r$, i.e. $\rm{Im}\{\delta k^{\sigma}_{\alpha,\theta} \}>0$. 
We find that the first pair, 
%$\bm{c}^{+}_{+,-} =\bm{c}^{-}_{+,+}$, 
$\bm{c}^{+}_{-,-} =\bm{c}^{-}_{-,+}$, satisfies these conditions,  provided that 
\begin{subequations}
\label{eq:transitions_zeromu}
\begin{equation}
B_{x0}< 2\Delta+B_{y0}, \quad B_{x0}+B_{y0}>2\Delta.
\end{equation}  
However,  addition requirement to the above inequalities is
\begin{equation}
B_{y0}<2\Delta+B_{x0}.
\end{equation}  
\end{subequations}
Otherwise, another solution with zero energy develops near $r=0$, formed by the pair $\bm{c}^{-}_{+,-} = \bm{c}^{+}_{+,+}$ in Eq.~\eqref{eq:4vecs}.
%%%The above analysis was carried for $B_{x0}\geq B_{y0}$.  In the opposite case, $B_{x0}< B_{y0}$, we can reduce the problem to the above solution by relabeling axes and changing the sign in front of $\sigma_x$ in Eqs.~\eqref{eq:Hchiral} and find that the allowed solution has the structure of the 4-component vector given by the pair .  
Overall, a non-degenerate solution of Eq.~\eqref{eq:Hchiral} with eigenenergy $E=0$ and localized near $r=0$ can exists within the rectangular region shown by bold solid lines in Fig.~\ref{fig:phase_diagram_mu}.  Since the energy gap vanishes on these lines, we identify the region inside as the topologically non-trivial superconducting phase that supports the MBS.  The outside region is the topologically trivial superconducting phase. 

We comment on the localization length of the MBS.  The length is determined by ${\rm max}\{1/|\delta k^{\sigma}_{\alpha,\theta}|\}$.  The localization length diverges near the phase boundaries, but then saturates to $\hbar v_F/\Delta$ in the center of the topological superconducting phase at $B_{x0}=B_{y0}$.  The predictions yielded by this continuum theory
for the dependence of the localization length on model
parameters will be compared with numerical
results in the next section.
\\

\section{Phase Diagram: Numerical Results}
\label{sec:Numerical}
\subsection{Discretized Hamiltonian}
\label{sec:DiscretizedHamiltonian}
We now show the results of numerical calculations in which the approximations that enable the analytical calculations in Sec.~\ref{sec:AnalyticalConsideration} are not made, obtaining results similar to those of the analytic model, as shown in
Fig.~\ref{fig:phase_diagram_mu}. We now consider a finite-length wire. We calculate the eigenvalues and eigenstates of a discretized version of Eq.~\eqref{eq:MainHamiltonian_integral} to determine the energy gap and identify the MBS. 
We rewrite the Hamiltonian representing the second derivative as a finite difference of the wave function $\bm{\Psi}_n = \bm{\Psi}(n\delta r)$ for a set of $n$ points separated by a discretization distance $\delta r$ along the wire; the total number of sites along the wire is $N=L/\delta r$.  The full Hamiltonian is given by the $4N\times 4N$ matrix 
\mgvcomment{I changed the text around Equations (22)-(26)!}
\begin{widetext}
\begin{equation}
\label{eq:H_discr}
\tilde H =\left(\begin{array}{ccccccc}
&&& \dots & & &\\
&\hat {\cal T} & \hat {\cal K}_{n-1} &\hat {\cal T} & \hat 0 & \hat 0& \\
\dots &\hat  0 &\hat  {\cal T} &\hat  {\cal K}_{n} &\hat {\cal T} &\hat 0&\dots \\
&\hat 0& \hat 0&\hat  {\cal T} &\hat  {\cal K}_{n+1} &\hat {\cal T} & \\
&&&\dots &&&
\end{array}\right),
\end{equation}
where the diagonal blocks are 
\begin{equation}
\label{eq:K}
\hat {\cal K}_n = \left(
\begin{array}{cccc}
2\tilde{T}-\mu+B_{z}^{n} & B_{x}^{n}-i B_{y}^{n} & 0 & \Delta  \\ 
B_{x}^{n}+i B_{y}^{n} & 2\tilde{T}-\mu-B_{z}^{n} & -\Delta 
&0 \\
0 & -\Delta & -2\tilde{T}+\mu-B_{z}^{n} & -B_{x}^{n}-i B_{y}^{n}\\
\Delta & 0& -B_{x}^{n}+iB_{y}^{n} & -2\tilde{T}+\mu+B_{z}^{n}
\end{array}\right),
\end{equation}
\end{widetext}
and the off-diagonal blocks are
\begin{equation}
\hat  {\cal T} = \left(
\begin{matrix}
-\tilde T &0&0&0 \\
0 &-\tilde T&0&0 \\
0 &0& \tilde T&0 \\
0 &0&0&\tilde T
\end{matrix}\right).
\end{equation}
The $\tilde{T}$ terms are given by 
\begin{equation}
\tilde{T}=\frac{\hbar^2}{2m\delta r^2}
\end{equation}
and originate from the discretized kinetic energy 
\begin{eqnarray}
-\frac{\hbar^2}{2m} \frac{\partial^2 \bm{\Psi}(r)}{\partial r^2} \to - \tilde{T}
\left(\bm{\Psi}_{n+1}-2\bm{\Psi}_n+\bm{\Psi}_{n-1}\right).
\end{eqnarray}
The $B^n_{x,y,z}$ terms in Eq.~\eqref{eq:K} are components of the magnetic field $\bm{B}^n=\bm{B}(n\delta r)$ at site $n$.
To be specific, we assume that the wire length $L$ is a multiple  of the magnetic period $\Lambda$.
We implement the boundary conditions $\Psi(r=0)=\Psi(r=L)=0$.

We diagonalize the discretized Hamiltonian and obtain the energy eigenvalues and eigenstates. The MBS, if present in the superconducting nanowire, is a non-degenerate state with energy in the middle of the superconducting gap that is zero in the limit of an infinite length wire and that is spatially localized at the ends of the nanowire.  We use these conditions to build phase diagrams for our setup for different amplitudes of the magnetic field components.  The eigenstates of the discretized Hamiltonian Eq.~\eqref{eq:H_discr} are $4N$ vectors, where 4 elements in each of $N$ blocks represent the four components of the electron wavefunction in the Nambu space.

We discuss the precise criteria for how we define zero energy and the gap energy in the numerical calculation for a finite system and the definition of the localization length in the following subsections of this section.

% figure with four panels
% I made composite figure by putting together 4 panels using Powerpoint
% four files are spectrum_atzerocomponent.pdf, spectrum.pdf, spectrum_fromspecialpoint.pdf, spectrum_fromspecialpoint_log.pdf
% figures need to be redrawn so the (a), (b), (c), (d) labels are consistent and also so parameters are as consistent as possible.

\subsection{Energy gap and zero-energy excitations}
\label{subsec:Results}

Using the numerical method described above, we analyze the low energy eigenstates of the Hamiltonian Eq.~\eqref{eq:MainHamiltonian_integral}.  
For the results shown, the discretization length used in the numerical calculations is $\delta r = 20$~nm
unless stated otherwise.
This value of $\delta r$ satisfies $\delta r \ll \Lambda$ and $\delta r \ll \hbar v_F/\Delta$, and we have checked
that changing the value of $\delta r$ does not change the numerical results significantly.
%To take a closer look at the evolution of the gap and formation of the zero-energy state,  w

%\snc{W}e can calculate the full energy spectrum  for the interval of magnetic fields $B_{x0}$ and $B_{y0}$ between $0$ and a few $\Delta$.
%, see Appendix B\snc{?check appendix label?}.  
To identify the phase transition and the development of the MBS, it is sufficient to focus only on the behavior of the lowest energy excitations.
To illustrate how MBS are manifest in the numerical results, we compare the lowest energy excitations for the ideal helical field with $B_{x0}=B_{y0}$ (where it is known that MBS are supported~\cite{klinovaja:prb12}) to the lowest energy excitations when one of the field components is zero (when there is no field chirality and the phase is topologically trivial for
all magnitudes of the nonzero component).

To explore the phase diagram in $B_{x0}-B_{y0}$ plane, we construct a color contour plot for $E_0$ for the wire of length $L=20~\mu$m, shown in Fig.~\ref{fig:phase_diagram_mu}.
%\snccomment{CHECK THAT THIS IS THE CORRECT FIGURE.} 
In the uncolored regions of the plot, the lowest energy $E_0$ is above $0.1\Delta=0.5~\mu$eV, which we identify as the gapped non-topological superconducting phase.    In wires of finite length, the zero energy state develops over finite crossover region shown as transition colors when $E_0<0.5~\mu$eV, where $E_0$ quickly drops below $10^{-2}\mu$eV. This region can be identified as the topological superconducting phase with a superconducting gap in the density of states and a low-energy state inside the gap, corresponding to the MBS.
We note that the crossover region is well described by the analytical expressions for the phase boundary evaluated in the previous section, see solid thick lines in Fig.~\ref{fig:phase_diagram_mu} and Eqs.~\eqref{eq:phase_boundary_analytical}.

\begin{figure*}[tb]
	%	\begin{center}
	\centering
	\includegraphics[width=0.33\linewidth]{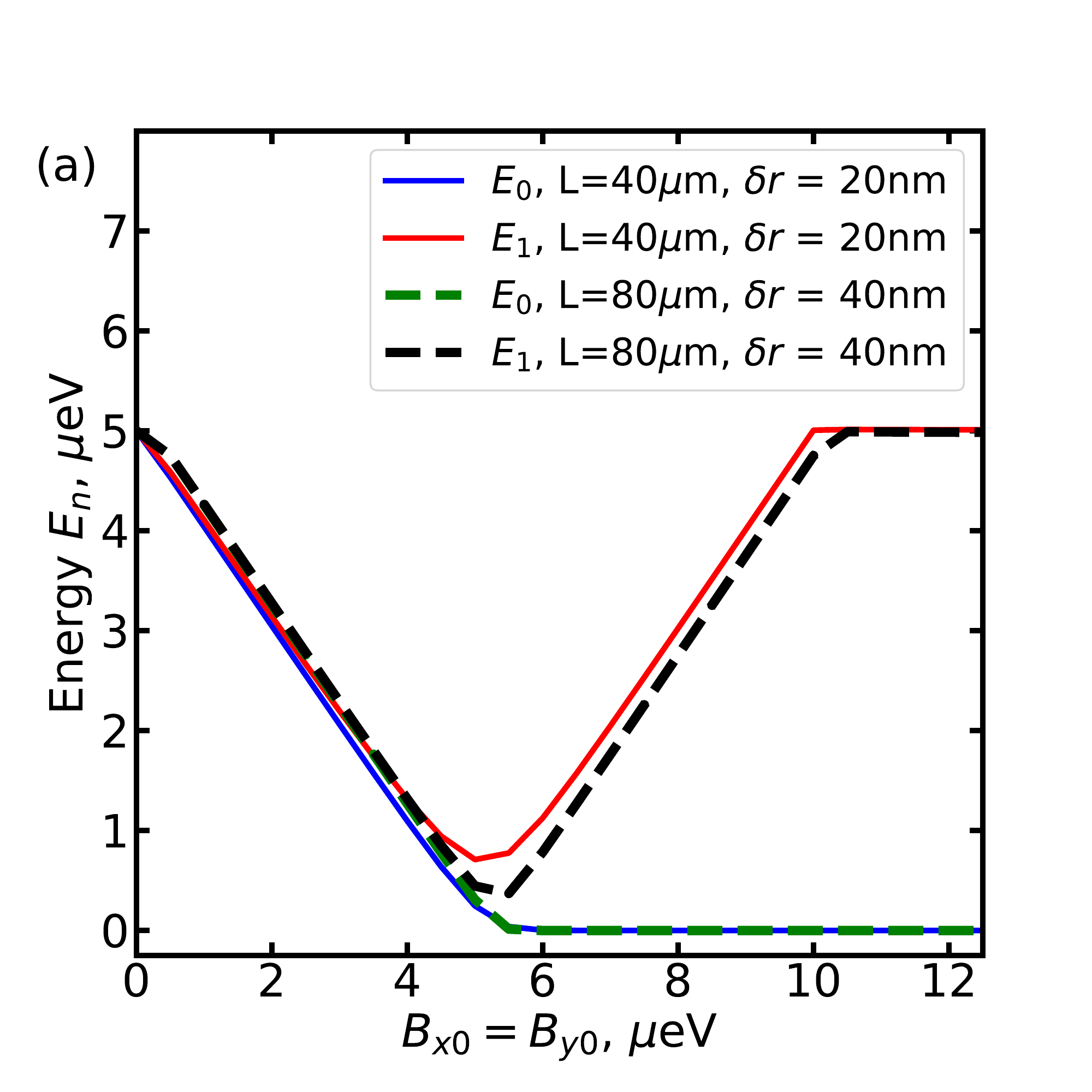}
	\includegraphics[width=0.33\linewidth]{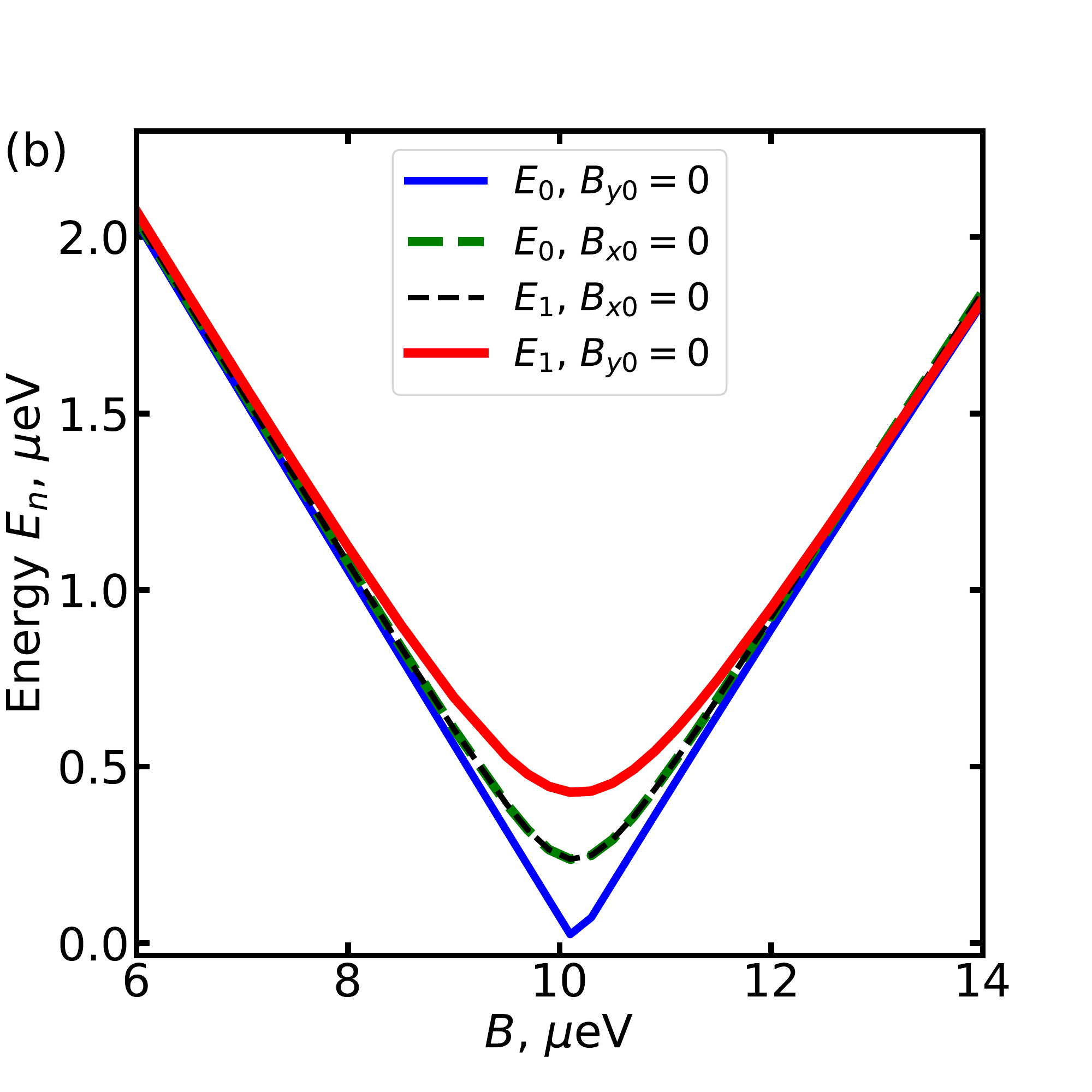}
	\includegraphics[width=0.33\linewidth]{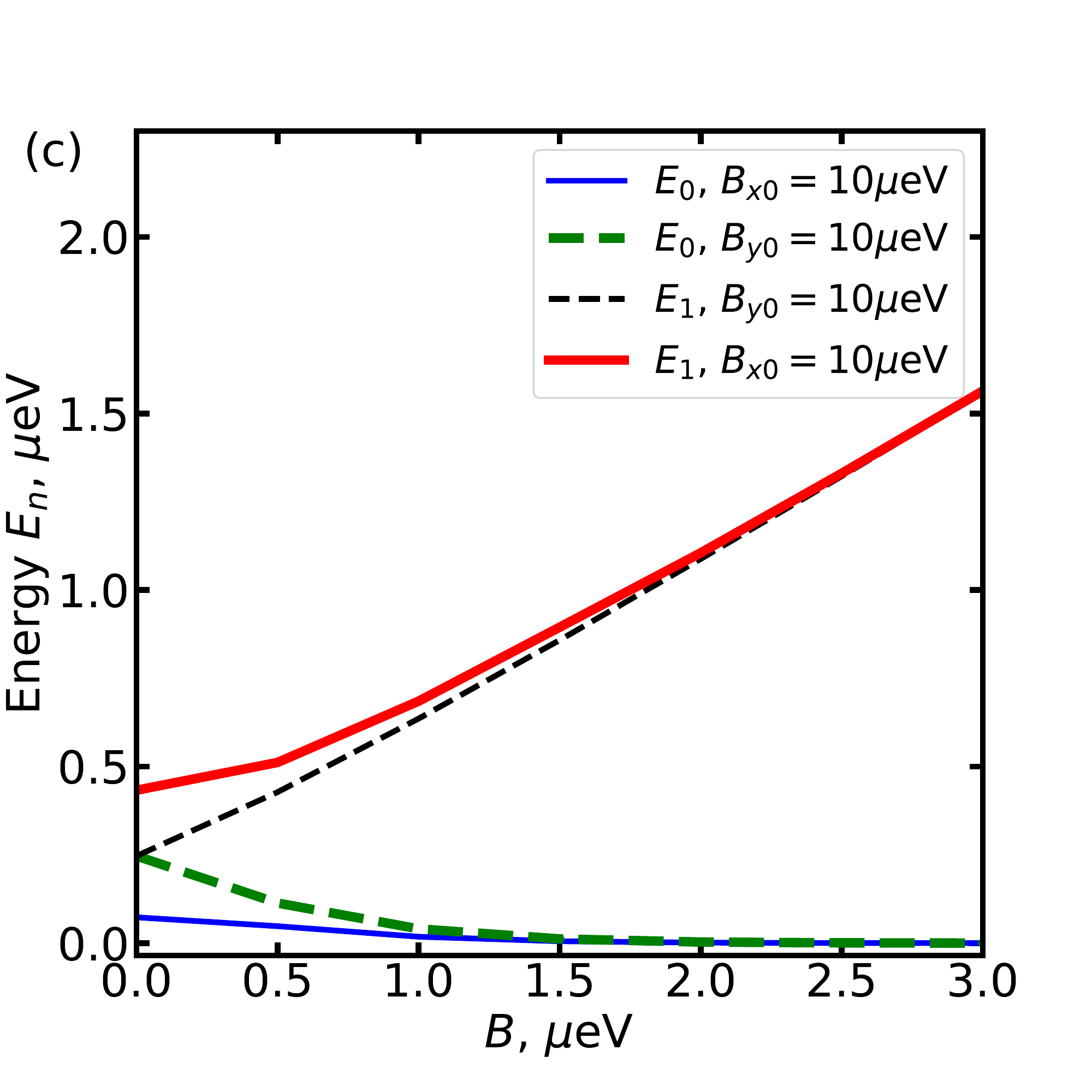}
	%\vspace{-6.3cm}
	\caption{
		%\snccomment{I rewrote the captions removing all mention of $\delta r$ in the figure captions unless it is different than $20$~nm; I added a sentence to the text giving that value and saying that it is used unless stated otherwise and is small enough that the results do not depend on it.} 
		Comparison of two lowest-energy excitation energies $E_0$  and $E_1$ in non-topological and topological phases as a function of magnetic field magnitude.
		(a): Energies $E_0$  and $E_1$ versus the magnetic field amplitude $B_{x0}=B_{y0}$ in wires evaluated for different wire lengths, solid lines: $L=40~\mu$m and $\delta r=20$nm; dashed lines: $L=80~\mu$m and $\delta r=40$nm.  
		%\snccomment{I would make the dashes wider so that the reader can see that $E_0$ for the shorter L at large B is also indistinguishable from zero.}
		%\mgvcomment{I cannot quickly run simulation for 80um and dr = 20 nm. To match caption this panel would need to have 2 curves removed and one curve redrawn with delta r = 20 nm (so that it has the same value for all the graphs).} 
		In the topologically nontrivial phase, $E_0$ approaches zero for when $B_{x0} = B_{y0}$ is large. 
		The second excitation energy $E_1$ reaches its minimum at the phase transition, where the gap in the density of states closes in the infinitely long wire, and then increases as the magnetic field is increased further.
		(b): Energies of the lowest two excitations as a function of one component of the field, $B_{x0}=B$ (solid lines) or $B_{y0}=B$ (dashed lines), \snccomment{mv- agree: !!changed this from 0: please check!!  Also, I would show the same number of eigenvalues in panels (a)-(c).,} while the other component is strictly zero, $B_{y0}=0$ or $B_{x0}=0$, respectively.  
		When one field component is zero, the magnetic field has no helicity and no topologically protected state can form. All the excitation energies show minima at $B\simeq 2\Delta=10~\mu$eV and then all energies increase as $B$ is increased further.  
		When $B_{x0}=0$ and the magnetic field is zero at the ends of the wire, $E_{0}=E_1$,  but the energies $E_{0}$ and $E_1$ are split near $B_{x0}=2\Delta$ when the field at the ends is not zero ($B_{y0}=0$ case).   
		(c): Energies of the lowest energy excitations as a function of one component of the magnetic field when the other component is fixed at twice the superconducting gap $\Delta$, either $B_{x0}=B,~B_{y0}=2\Delta$ or $B_{y0}=B,~B_{x0}=2\Delta$.
		For both cases, the energy of the lowest-energy state monotonically decreases towards zero as $B$ increases, while all other excitation energies increase as the superconducting gap opens, consistent with a topologically nontrivial phase.
		For all these plots,  the proximity-induced superconducting order parameter $\Delta = 5~\mu$eV, and the chemical potential  
		$\mu=\hbar^2Q^2/8m = 49.6~\mu$eV is matched to the middle of the magnetic superlattice gap.
		%\mgvcomment{remove:for a perfectly helical field.  
		For panels (b) and (c), the wire length is $L=40~\mu$m.}
	%\snccomment{Is this statement about the chemical potential correct?}
	\label{fig:low_energies}
	%	\end{center}
\end{figure*}

\mgvcomment{Text is adjusted to Fig. 4}
To illustrate the actual dependence of lower energies on the magnetic field, we show the two lowest two energies $E_{0,1}$ as function of the magnetic field strength $B_{x0}=B_{y0}=B_0$ in Fig.~\ref{fig:low_energies}(a), this field configuration corresponds to a perfect helix studied earlier\cite{braunecker:prb10,klinovaja:prb12,egger:prb12}.
When $B_0=0$, the values of both energies are just above the superconducting gap $\Delta$.
As $B_0$ is increased from zero, the effective superconducting gap $\left | B_0-\Delta \right |$ as well as all three eigenenergies decrease.   As $B_0$ is increased beyond $\Delta$, the lowest energy $E_0$ continues to decrease monotonically towards zero, while the energy of the higher eigenstate goes through its minimum at $B_{0}\simeq\Delta$ and then increases until it reaches an asymptotic value equal to the superconducting gap $\Delta$. 
Here, there is a topologically nontrivial phase when $B_{x0} = B_{y0} > 2\Delta$, and the lowest energy $E_0$ approaches zero while the higher energies increase as strength of the field $B_{x0} = B_{y0}$ are increased past $2\Delta$.

The form of the phase diagram shown in Fig.~\ref{fig:phase_diagram_mu} makes it clear that the robustness of the MBS to eccentricity of the magnetic field helicity depends strongly on the magnitude of the larger field component; in fact, the topological phase could be reached even when one of the magnetic field components is much smaller than the other, provided that the larger component is near $2\Delta$.  To  explore this region of the phase diagram in more detail, we plot the two lowest energy states, $E_0$ and $E_1$ as a function of one component of the magnetic field, $B_{x0}$ or $B_{y0}$, while keeping the other component equal to zero, see Fig.~\ref{fig:low_energies}(b).
We observe that the lowest energy $E_0$ reaches its minimum when the non-zero component is $\approx 2\Delta$ and then increases as the field component is increased further.  We note that in these plots, for which one component of the field is zero, there is no topologically protected phase.

Figure~\ref{fig:low_energies}(c) shows the energies of the two lowest-energy excitations, $E_{0,1}$ as function of the magnetic field strength of one component, while the other is fixed at $2\Delta$. We notice that for both $B_{x0}=2\Delta$ (solid lines) or $B_{y0}=2\Delta$ (dashed lines), the lowest energy excitation vanishes quickly as the magnitude of the other field component is increased.
\snccomment{Fig.~\ref{fig:low_energies}(d) shows the same numerical data as Fig.~\ref{fig:low_energies}(c) on a semilog scale,
	demonstrating that the energy decreases exponentially as the magnitude of the smaller field component is increased.}  
\snccomment{Can we delete the rest of this paragraph?
At the same time, all higher energy excitations monotonically increase with the magnitude of the variable magnetic field component.  The energy spectrum in the region of the phase diagram when one component is close to $2\Delta$ and the other is finite but small supports our previous observation that these regions can be identified as the topological superconducting phase holding the MBS.  }
%\mgvcomment{ from a different section:} 
%The minimal value of the \snc{energy of the state with second--lowest energy} 
%\snccomment{is this correct? I reworded because I think ``second excited state" is ambiguous} 
%corresponds to the energy quantization of electrons in the wire of length $L$ for electrons with momentum $\hbar Q/2$, while the other energy eigenstates increase with the variable field component as the superconducting gap opens.}
%

While the actual orientation of magnetic field components $B_{x0}$ and $B_{y0}$  is arbitrary with respect to the direction of the wire, the dependence of the energies on the two components are not identical since the field magnitude at the ends of the wire is determined by the value of $B_{x0}$, while the $B_y$ field always vanishes at the wire ends, see Eq.~\eqref{eq:BxBy}.  This distinction between components explains a weak asymmetry of the phase diagram with respect to line $B_{x0}=B_{y0}$ in Fig.~\ref{fig:phase_diagram_mu}.  The distinction is even more pronounced in the energy plots shown Figs.~\ref{fig:low_energies}(b) and \ref{fig:low_energies}(c).  When  $B_{x0}=0$ and $B_{y0}\neq 0$, the low energy levels, such as $E_{0,1}$, are doubly degenerate.  In the opposite case,  $B_{x0}\neq 0$ and $B_{y0}=0$, this double degeneracy is split, pushing the lowest energy closer to zero, see the lowest solid line in Fig.~\ref{fig:low_energies}(b).  The double degeneracy is always split by $B_{x0}\neq 0$, as shown by the lower two dashed lines in Fig.~\ref{fig:low_energies}(c).  

Overall, the above analysis demonstrates that the topological phase with MBS is possible when the dominant magnetic field has magnitude about $2\Delta$ and the minor component is strong enough to open the gap in the energy spectrum and push the energy of the MBS to zero.  %The optimal regime is further achieved 
MBS are enhanced further when the dominant field component is at its maximum value at the wire ends.  
	
%\begin{figure}[tb]
%%	\begin{center}
%	\centering
%		\includegraphics[width=0.99\linewidth]{ground_energy.pdf}
%		\caption{
%			The lowest positive energy $E_0$ as function of the magnetic field amplitude $B_{x0}=B_{y0}$ in wires evaluated for  different choices of  \snc{wire length} $L$ and \snc{discretization length} $\delta r$\snc{.  Solid black line}: $L=80~\mu$m and $\delta r=40$~nm, \snc{red dashed line}: $L=40~\mu$m and $\delta r=20$~nm, \snc{blue dash-dotted line}: $L=40~\mu$m and $\delta r=40$~nm, \snc{green dotted line}: $L=20~\mu$m and $\delta r=20$~nm. The proximity-induced  superconducting order parameter $\Delta = 5~\mu$eV and the chemical potential  $\mu=\hbar^2Q^2/8m = 49.6~\mu$eV is matched to the middle of the magnetic superlattice gap. The onset of exponential decrease of $E_0$ with increasing $B_{x0}=B_{y0}$ indicates the transition to the topological superconducting phase.
%		}
%		\label{fig:E0}
%%	\end{center}
%\end{figure}

% Replace Fig 5 and Fig 6 with a composite figure showing two ways to calculate correlation length
\begin{figure*}[tb]
	\centering
%		\includegraphics[width=0.99\linewidth]{localization_figure_composite_17_september.png}
%		\vspace{-.5cm}
\includegraphics[width=0.45\linewidth]{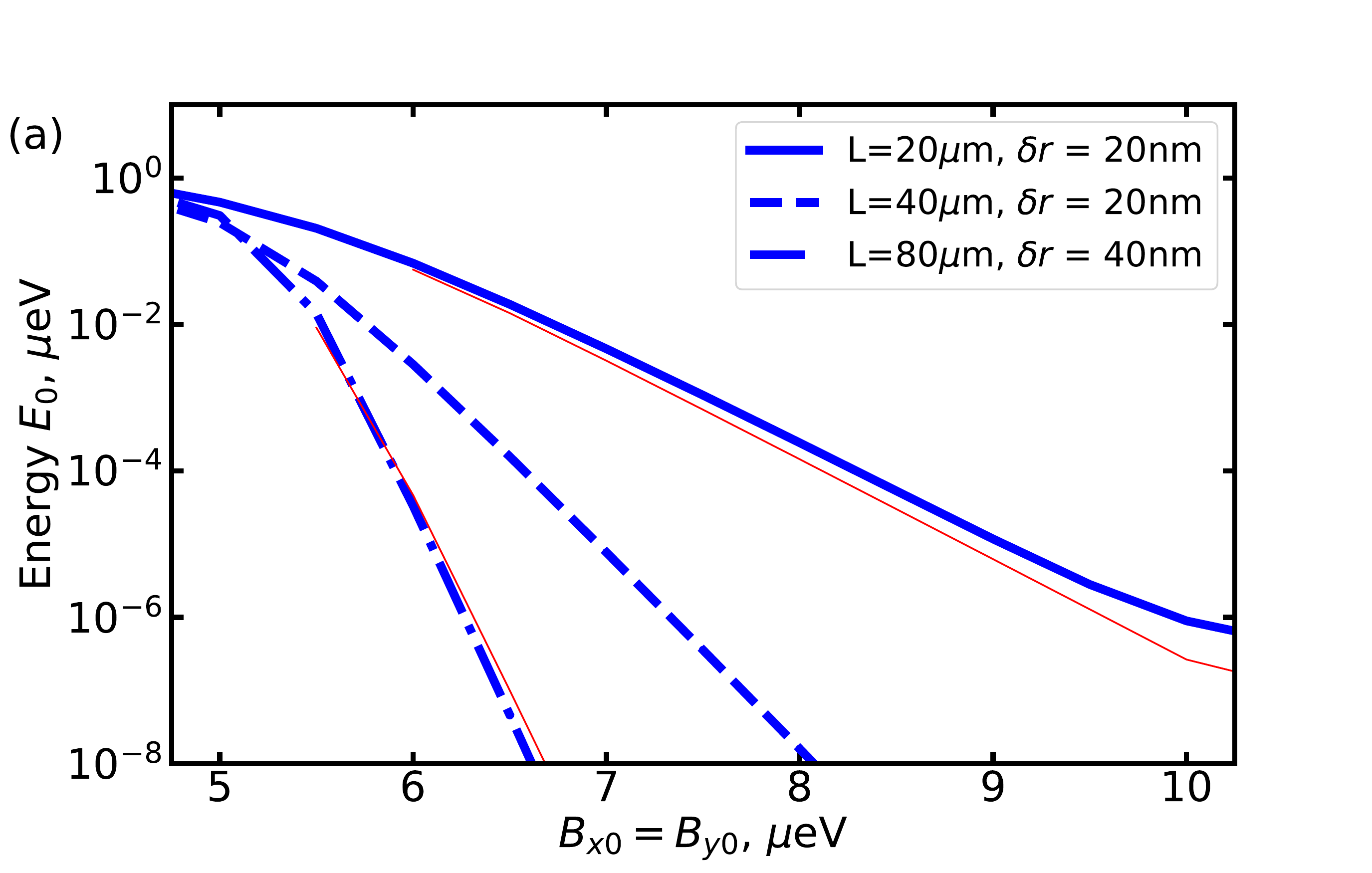}
\includegraphics[width=0.45\linewidth]{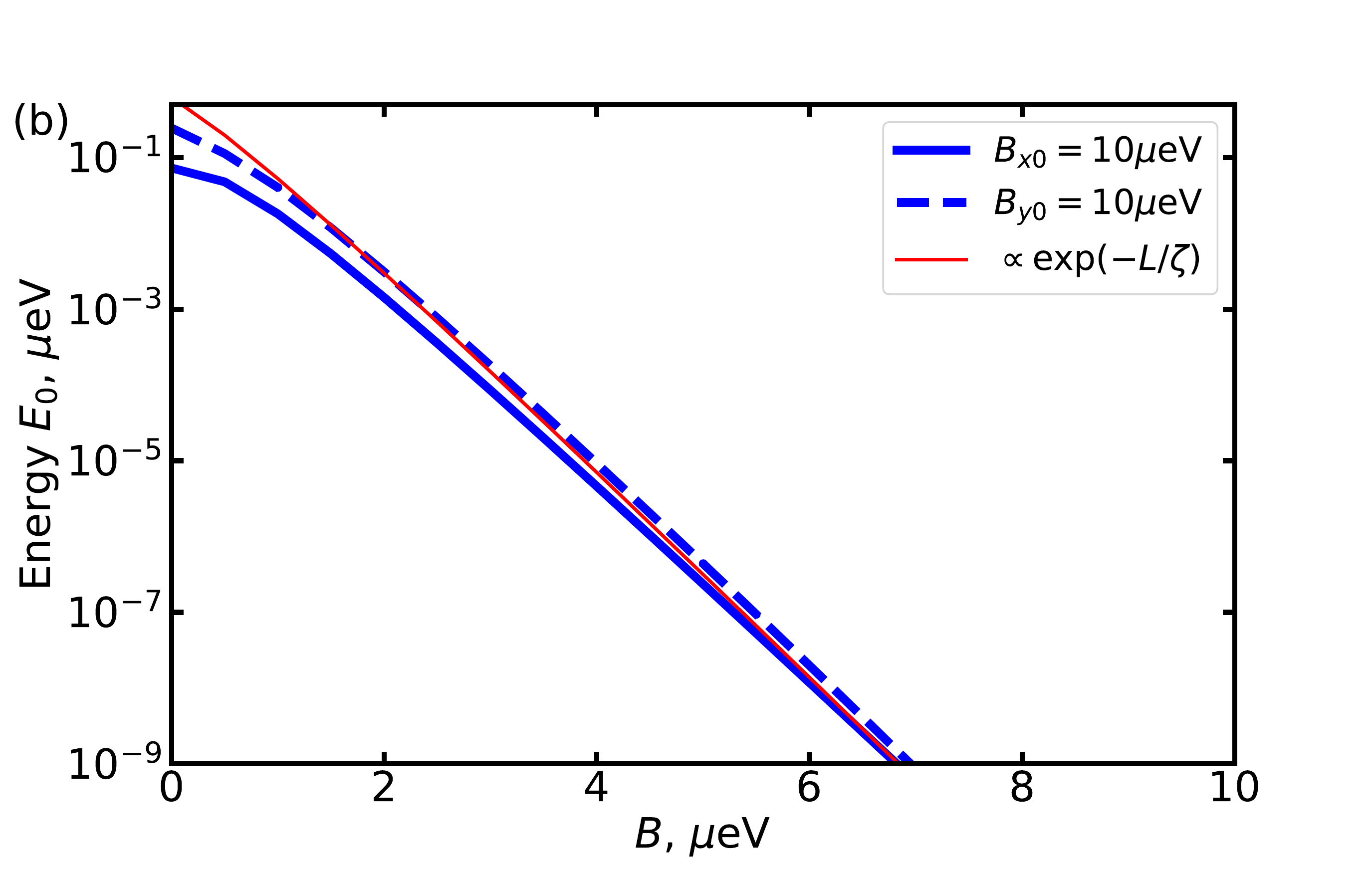}
\includegraphics[width=0.45\linewidth]{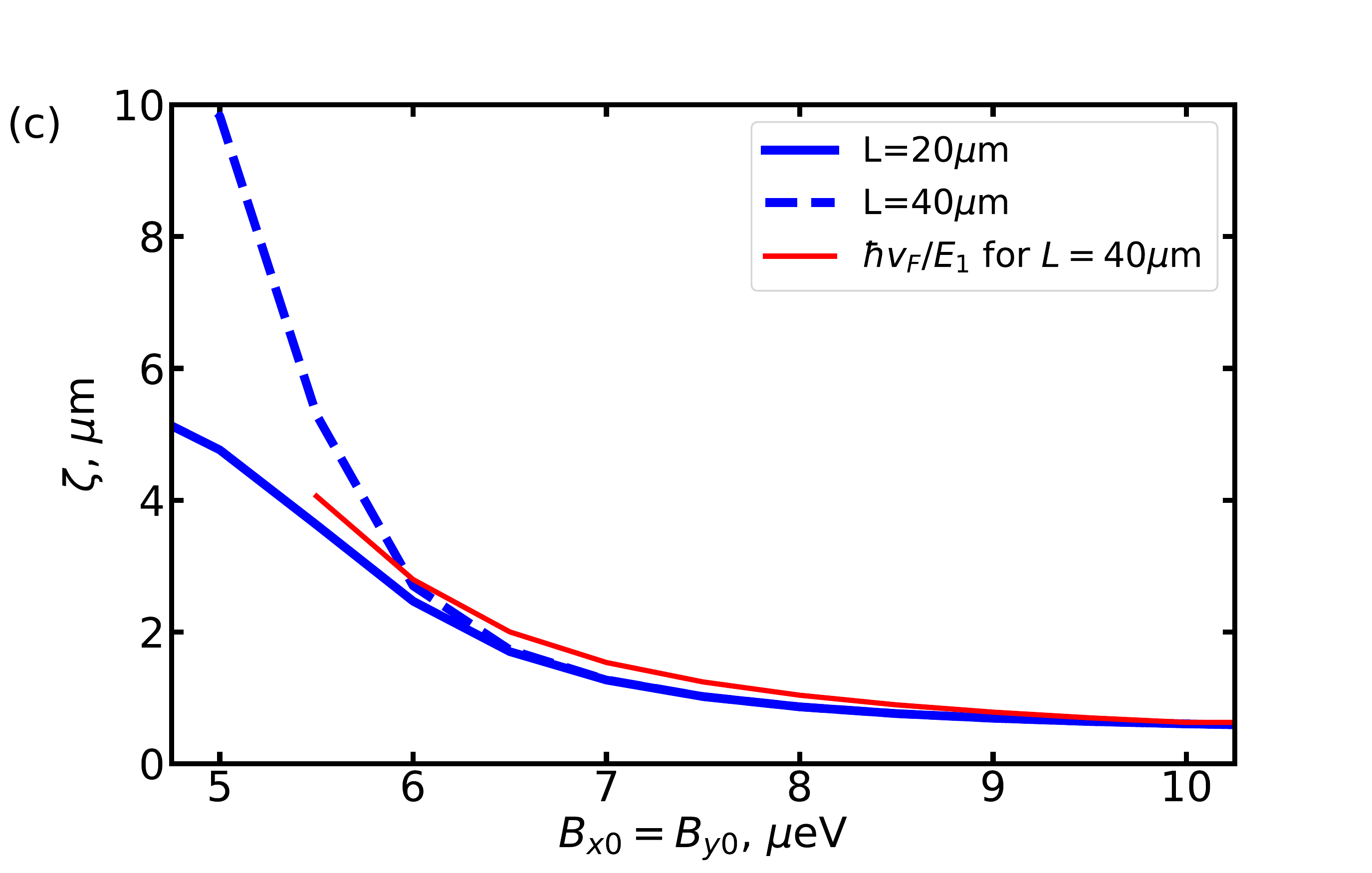}
\includegraphics[width=0.45\linewidth]{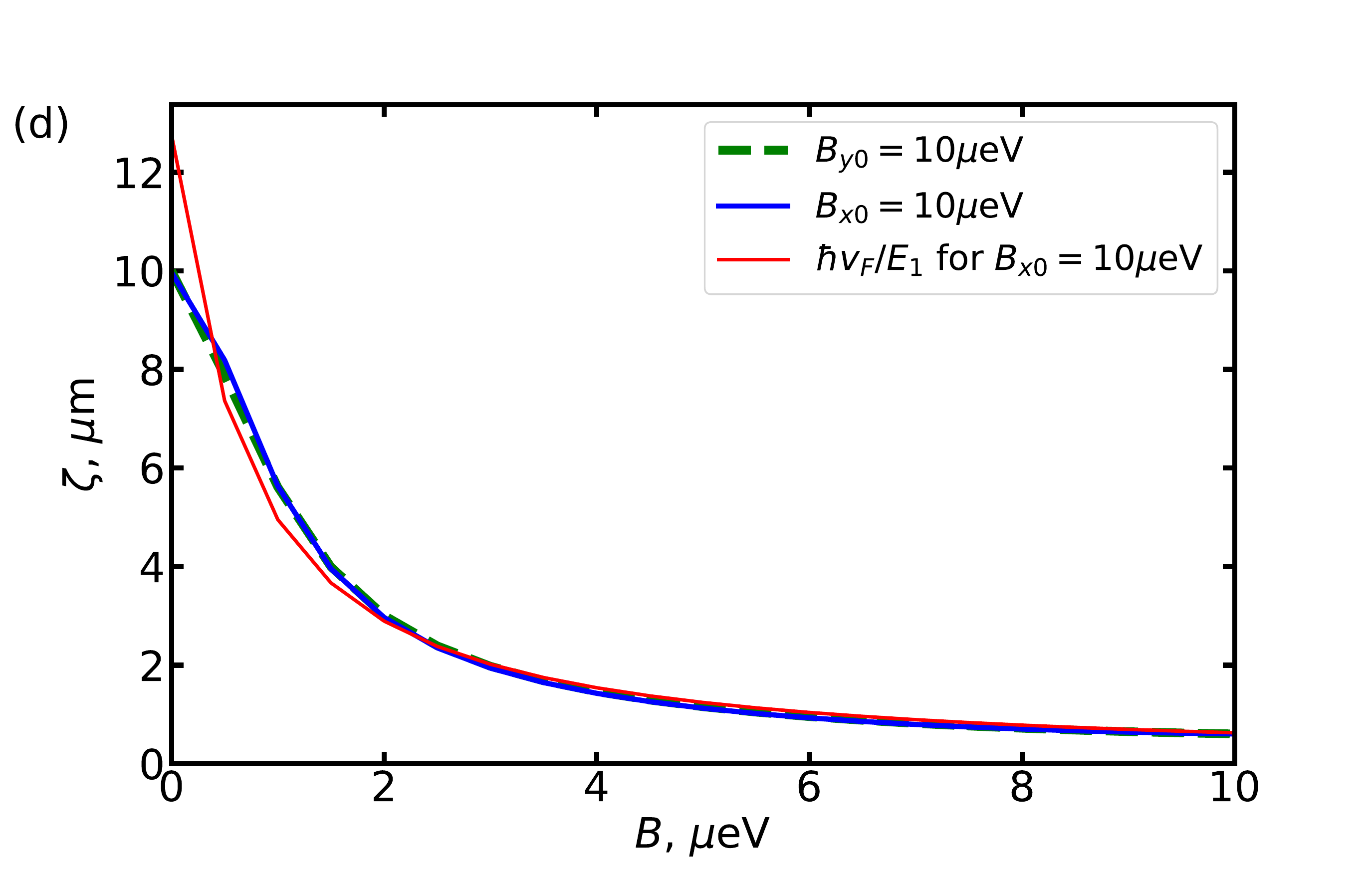}
\caption{
(a): Numerically obtained value of the lowest positive energy $E_0$ as function of the magnetic field amplitude $B_{x0}=B_{y0}$ in wires of length 
$L=20\mu$m (solid line) and $L=40\mu$m (dashed line) obtained using $\delta r =20$nm and $L=80\mu$m	obtained using $\delta r =40$nm.
%%% $L=80$m (solid black line), $40$, (red dashed line), and $20~\mu$m (green dotted line), with  
The onset of exponential decrease of $E_0$ with increasing $B_{x0}=B_{y0}$ above $\Delta$ indicates the transition to the topologically nontrivial superconducting phase, with the stronger dependence in longer wires arising because the overlap between MBS decreases exponentially with wire length.
Thin red lines show $E\propto \exp(-L/\xi)$, where $\xi$ is given by Eq.~\ref{eq:xi_E1}.
(b): Numerically obtained value of the lowest positive energy $E_0$ in wires of length $L=40\mu$m as function of the magnetic field amplitude $B_{y0}$ 
(solid line)  or $B_{x0}$ (dashed line) fixed with $B_{x0}=2\Delta$ or  $B_{y0}=2\Delta$, respectively.  The similarity of the behavior to that seen in (a) demonstrates the robustness of the topologically nontrivial phase to ellipticity in the helical magnetic field.  
Energy $E_0$ decreases exponentially as $L/\xi$ is increased, as demonstrated by thin red line for $\propto \exp(-L/\xi)$, where $\xi$ is computed from Eq.~\eqref{eq:xi_E1} using $E_1(B_{y0})$ at $B_{x0}=2\Delta$, see red solid line in  Fig.~\ref{fig:low_energies}(c).
(c): Numerical obtained localization length of the lowest-energy state $\zeta$, obtained using Eq.~\eqref{eq:loc_length} versus the strength of the magnetic field for a perfect helix, $B_{x0}=B_{y0}$, in wires of length $L= 20$ (solid  line), $40~\mu$m (dashed line). Also shown as the thin red line is the localization length $\xi$, Eq.~\eqref{eq:xi_E1}, obtained using values of $E_1$  for a wire of length $L=40~\mu$m, see red solid line in Fig.~\ref{fig:low_energies}(a). 
%%%\mgvcomment{Remove -- too computationally expansive -- by fitting the dependence of the energy $E_0$ as a function of wire length $L$ in (a) to the functional form $E_0 = C e^{-2L/\zeta}$, with $C= ?$ obtained from the best fit for $B_{x0}>6~\mu$eV.}
%
(d): Numerically obtained localization lengths of the lowest-energy state, obtained using Eq.~\eqref{eq:loc_length} versus the magnitude of $B_{x0}$ (dashed line) or $B_{y0}$ (solid line) with $B_{y0}=2\Delta$ or $B_{x0}=2\Delta$, respectively, in wires of length $L=40~\mu$m. Also shown as the thin red line  is the localization length $\xi$ obtained from Eq.~\eqref{eq:xi_E1} using $E_1(B_{x0}=2\Delta,B_{y0}=B)$ as function of $B$, see solid red line in Fig.~\ref{fig:low_energies}(b). The similarity to (c) demonstrates the robustness of the topological phase to ellipticity in the helical magnetic field.
In all panels, superconducting gap parameter $\Delta = 5~\mu$eV and chemical potential  $\mu=\hbar^2Q^2/8m = 49.6~\mu$eV, which is matched to the middle of the magnetic superlattice gap.} % end of caption
		\label{fig:localization_length_figure}
%	\end{center}
\end{figure*}

\subsection{Localization length}

\mgvcomment{Some significant changes until Eq. 29}
The coherence length is important energy scale of a superconductor and is inversely proportional to the superconducting energy gap:
\begin{equation}
\label{eq:xi_E1}
\xi = \frac{\hbar v_F}{E_1}.
\end{equation}
Equation~\eqref{eq:xi_E1} takes into account that in the topological superconducting phase, the lowest energy state corresponds to the MBS and the superconducting gap is determined by  the next positive energy $E_1$.
In this subsection we argue that the localization length  $\zeta$  of the MBS near the wire ends is consistent with the correlation length determined by Eq.~\eqref{eq:xi_E1}, $\zeta\simeq \xi$.
We also show that the lowest positive energy $E_0$  agrees well with the exponential dependence on wire length $L$ as $E_0 \propto \exp(-L/\xi)$.
The behavior of the localization length for helical fields with elliptical cross-section is qualitatively similar to that found
for purely helical fields.

Figure~\ref{fig:localization_length_figure}(a) shows on a semilog scale the lowest energy as a function of $B_{x0}=B_{y0}$, corresponding to a perfectly helical field.~\cite{klinovaja:prb12}  For $B_{x0}=B_{y0}=B_0>\Delta$ the lowest excitation eigenenergy $E_0>0$ decreases exponentially with $B_0$, as demonstrated in Fig.~\ref{fig:localization_length_figure}(a).
To demonstrate the dependence on wire length, we show the energy versus $B_0$ for wires with length $L=20$ (solid line), $L=40$, (dashed line) and $L=80~\mu$m (dash-dotted line) and compare the result with exponential fit $\propto \exp(-L/\xi)$, where $\xi$ is evaluated 
from Eq.~\eqref{eq:xi_E1} with numerical values of $E_1$ ($E_1$ is presented in Fig.~\ref{fig:low_energies}(a) for $L=40$ and $80\mu$m.)
%Notice that the \snc{energies at the} minima decrease as $L$ increases.  
%\snccomment{This semilog plot shows the dependence of the lowest energy $E_0$  versus $B_0$ on a semilog scale for $L=20, 40,$ and $80~\mu$m. $E_0$ decays exponentially with $L$, which is consistent with $E_0$ being due to overlap between two states that are localized on the opposite wire ends.}  

Figure~\ref{fig:localization_length_figure}(b) examines the case of an elliptical helical field; it is a semilog plot of the three lowest excitation energies as a function of one field component (either $B_{x0}$ or $B_{y0}$) as the other is held fixed at $2\Delta$.
As the variable magnetic field component is increased from zero, the lowest energy $E_0$ decreases towards zero.
The lowest energy $E_0$ again decreases exponentially with the wire length $L$ as $\propto \exp(-L/\xi)$ with $\xi$ evaluated from Eq.~\eqref{eq:xi_E1} with $E_1$ shown in Fig.~\ref{fig:low_energies}(c) by a dashed line for $B_{x0}=2\Delta$ and variable $B_{y0}$.

%\snccomment{I think at this point we should compare the analytic prediction for the localization length to the numerical results.}

%\snccomment{I propose to remove this discussion of the delta r dependence in figure 5 because I stated that the results don't depend on them above.  I would like to incorporate Fig.\ 5 as a panel with Fig.\ 6 and use it to illustrate that the exponential dependence of $E_0$ on $B$ can be interpreted as overlap between the Majoranas, and then extract a correlation length that can be compared to the other method of calculating the correlation length that is shown in Fig. 6.  If we do the latter, then I would propose plotting for the same case as Fig. 6 (By0 fixed, vary Bx0) and extract a correlation length by fitting an exponential to the dependence of E0 on L, and comparing the result for a few values of Bx0 to the curves plotted in Fig. 6.}
%To \snc{check that the numerical value of the lowest energy is not sensitive to the discretization length} $\delta r$, we present \snc{in Fig.~\ref{fig:E0}} two curves for wire length $L=40~\mu$m  evaluated for $\delta r= 20$ and $40$~nm.  We conclude that as long as $\delta r \ll \Lambda$ and $\delta r \ll \hbar v_F/\Delta${, the value of $\delta r$} does not significantly change the value of $E_0$.
%\snccomment{End of text that I'd like to delete.}

We also calculate the localization length by examining the wavefunction of the lowest energy excitation
(the eigenvector with energy eigenvalue $E_0$).
To characterize the localization length, we define the following integral expression that effectively evaluates the distance of the `center-of-mass' of the MBS wave function from the wire ends:
\begin{equation}
\label{eq:loc_length}
\zeta = 2\delta r\left( 
\sum_{n=0}^{N/2}    nP_n  +
\sum_{n=N/2}^N  (N-n) P_n\right),
\end{equation}
where $P_n$  is the probability density for the MBS at site $n$:
\begin{equation}
P_n = \bm{\Phi}_n^\dagger \bm{\Phi}_n,\quad \bm{\Phi}_n = \bm{\Phi}(n\delta r), 
\end{equation}
and 
$\bm{\Phi(r)}$ is obtained from the state $\bm{\Psi(r)}$ corresponding to the lowest positive eigenenergy of Eq.~\eqref{eq:BdG} via transformation~\eqref{eq:MajoranaBasis}.

The dependence of the localization length $\zeta$ on the magnetic field is shown in Fig.~\ref{fig:localization_length_figure}(c)  a perfect helical field as a function of field magnitude and in Fig.~\ref{fig:localization_length_figure}(d) for the case where one field component is fixed at $2\Delta$ and the magnitude of the other component is varied.  We calculate the localization length using Eq.~\eqref{eq:loc_length} for several values of the nanowire length.  At $B_{x0}=B_{y0}<\Delta$, evaluated values of the localization length $\zeta $ is comparable to the length of the wire.  
%In this case the state is completely delocalized.  
At larger fields, $\zeta$ decreases rapidly and reaches the value $\zeta \simeq \xi/2$, where  $\xi_{\rm sc} = \hbar^2 Q/(2m\Delta)\simeq 1.26~\mu$m the superconducting coherence length that determines $|\Psi_n|^2\propto \exp(-2n\delta r/\xi)$.  
Figs.~\ref{fig:localization_length_figure}(c) and \ref{fig:localization_length_figure}(d) also show that the estimate of the localization
length using the dependence of the coherence length $\xi$ on $E_1$ agrees well with the estimate of the MBS localization length done using Eq.~\eqref{eq:loc_length}.
%\snccomment{I would like to take figure 5 (removing the different delta r curves) as a panel here and extracting a correlation length from the dependence of E0 on L for a few values of B, and then comparing that length to the one plotted in the current figure 6.  This could be done both for the perfect helix and for the elliptical helix.}  

%
%\begin{figure}[tb]
%\centering
%%\begin{center}
%\includegraphics[width=0.95\linewidth]{loc_length.pdf}
%\caption{Dependence of the localization length of the \snc{lowest}-energy state on the strength of magnetic field $B_{x0}=B_{y0}$ for the perfect helix in wires of different length and for $\delta r = 20$ and $40$~nm.  The proximity-induced superconducting order parameter $\Delta = 5\mu$eV and the chemical potential is  matched to the middle of the gap of the magnetic superlattice, $\mu =\hbar^2Q^2/8m = 49.6\mu$eV.  \snccomment{I think we should add a panel showing the localization length varying one component of field with the other fixed at 2 Delta.}
%}
%\label{fig:loc_length}
%%\end{center}
%\end{figure}

\subsection{Dependence of the phase diagram on chemical potential}

We now investigate the phase diagram when the chemical potential $\mu$ is not located in the middle of energy spectrum gap of magnetic superlattice, so that $\mu = \hbar^2Q^2/8m+\delta \mu$ with $\delta\mu\neq 0$.

It is straightforward to extend the analytic theory developed in Sec.~\ref{sec:AnalyticalConsideration} to the case
in which $\delta\mu \ne 0$.
The excitation energies of Hamiltonian~\eqref{eq:Hchiral} are 
%given by the following expressions
\begin{equation}
E_\theta^\pm = \sqrt{B_\theta^2+\Delta^2+\delta\mu^2+v_F^2\delta k^2\pm 2 D}~,
\end{equation}
with
\begin{equation}
D=\sqrt{B_\theta^2(\Delta^2+\delta\mu^2)+\delta\mu^2v_F^2\delta k^2}.
\end{equation}
The lowest energy for real $\delta k$ is achieved for $\delta k=0$ and is given by
\begin{equation}
E^-_\theta = |B_\theta - \sqrt{\Delta^2+\delta\mu^2}|,
\end{equation}
and the gap closes for $B_\theta=\sqrt{\Delta^2+\delta\mu^2}$.  This expression is similar to the condition for the point of the phase transition in perfect helical magnetic field\cite{braunecker:prb10,egger:prb12}.  For fixed $\Delta$ and $\delta\mu$, the gap closes when 
% the magnetic field reaches  
\begin{subequations}
\label{eq:transitions_nonzeromu}
\begin{eqnarray}
\label{eq:cond_Bp}
B_+ = (B_{x0} + B_{y0})/2 > \sqrt{\Delta^2+\delta\mu^2}.
\end{eqnarray} 
This condition specifies the magnitude of the magnetic field necessary to develop the topologically nontrivial superconducting phase.\cite{lutchyn:prl10, oreg:prl10, klinovaja:prl12}  The second condition for the topological phase is determined by ellipticity of the spiral magnetic field that limits the relative mismatch between $B_{x0}$ and $B_{y0}$ components.  At non-zero $\delta\mu$, the corresponding condition is 
\begin{equation}
\label{eq:transitions_nonzeromu_Bminus}
|B_-|< \sqrt{\Delta^2+\delta\mu^2}~,
\end{equation}  
\end{subequations} 
with $B_- = (B_{x0} - B_{y0})/2$.
We notice that Eq.~\eqref{eq:transitions_nonzeromu_Bminus} implies that non-zero $\delta\mu$ makes the system more robust to imperfect helical magnetic fields, but at the same time, Eq.~\eqref{eq:cond_Bp} implies that stronger fields are needed to reach the topologically nontrivial phase. 
The conditions for the existence of the MBS at $\delta \mu=0$ can be interpreted as requiring that one out of the two electron
bands has a gap in the energy interval
%the condition when only one out of two electron bands of the magnetic superlattice crosses with energy 
$\hbar^2Q^2/8m\pm\Delta$ while the other does not.
%, the latter expressions describe the energy of the excitations on the superconductor for particles with Fermi momentum.  
When $\delta \mu \ne 0$, we find that a similar condition applies.  If at the Fermi momentum one of the bands is gapped while the other is not, then a topologically nontrivial phase can be supported,
The mismatch of the Fermi momentum with the points of 1D Brillouin zone correspond to the superconducting excitation with energy $\hbar^2Q^2/8m\pm\sqrt{\Delta^2+\delta \mu^2}$, and the MBS exists if these energies cross one and only one band of electrons in the magnetic superlattice, 
see Fig.~\ref{fig:magnetic_bands}.

\begin{figure}[tb]
\begin{center}
\includegraphics[width=0.95\linewidth]{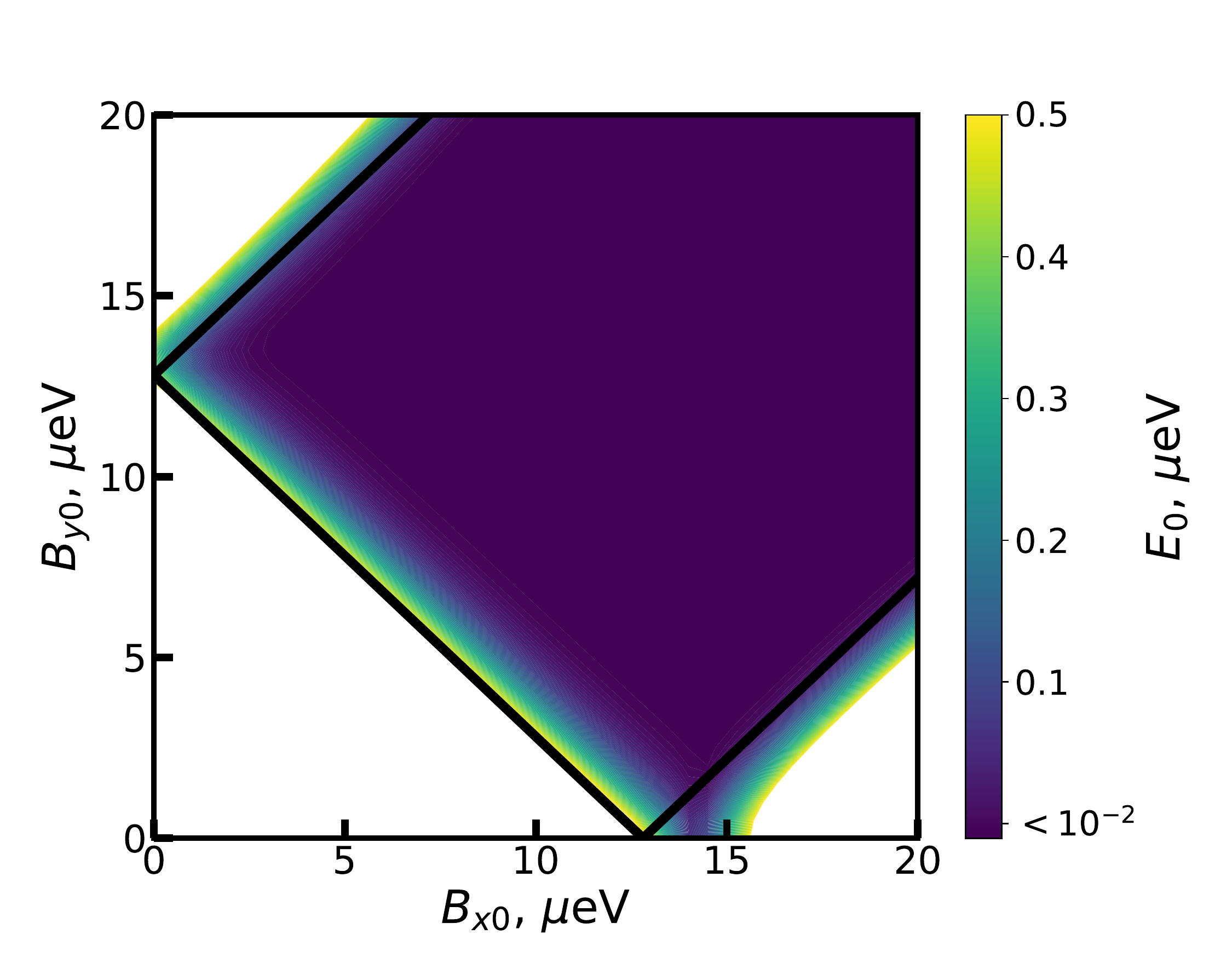}
\caption{Contour plot of the lowest positive energy $E_0$ of the system in the plane of magnetic field amplitudes $B_{x0}$ and $B_{y0}$ for the chemical potential $\mu=\hbar^2Q^2/8m+\delta \mu$ with $\delta \mu = 4~\mu$eV.   Here, the %proximity-induced  
superconducting pairing energy $\Delta = 5~\mu$eV, the and the wire length $L=20~\mu$m.   \snccomment{I assumed that the figure will have $\delta r = 20 nm$ and so didn't mention it.  I put in the question mark because if it is straightforward, it would be good to use a 20 micron wire so that the figure is directly comparable to Fig. 3.}
The region with energy $E_0$ above $0.5~\mu$eV corresponds to the gapped non-topological superconducting phase and is shown as unfilled parts of the $B_{x0}$--$B_{y0}$ plane.  The solid straight lines represent the analytical expressions \eqref{eq:transitions_nonzeromu} for the phase boundaries derived for linearized bands and an infinite length wire.  In wires of finite length, the zero energy state develops over finite crossover region shown as transition colors in the color plot for the region of $E_0<0.5~\mu$eV, the dark purple color shows area where $E_0$ drops below $10^{-2}\mu$eV.
%\snccomment{Do we know why the shift predicted analytically doesn't appear to be accurate?  Could it just be a finite-size effect?  If so, maybe redo for a longer wire?}
}
\label{fig:phase_diagrams_dmu4}
\end{center}
\end{figure}

We also performed numerical investigations of systems with $\delta \mu \ne 0$ for wires of finite length.
Figure~\ref{fig:phase_diagrams_dmu4} shows the results for a phase diagram obtained for a wire with length $L= 20~\mu$m by showing the energy of the lowest-energy state as a function of the magnitudes of the magnetic field components.  
We define the regions of field in which the system is in a topologically
nontrivial phase and MBS are supported to be those where the lowest-energy state has energy that is much less than that of the superconducting gap.
A comparison of the results in Fig.~\ref{fig:phase_diagrams_dmu4} with those in Fig.~\ref{fig:phase_diagram_mu} demonstrate that tuning the chemical potential of the nanowire can play an important role in optimizing the robustness of MBS.

\section{Conclusions}
\label{sec:conclusions}
Motivated by the possibility of introducing strong artificial spin-orbit coupling in nanowires fabricated in silicon, we
have considered a nanowire-superconductor hybrid structure with a non-uniform magnetic field and studied the conditions for which {the superconductivity is topologically nontrivial and MBS appear}. We have investigated both analytically and numerically the case of a spiral magnetic field with an elliptical cross-section, which becomes helical for a round cross-section. This spatial dependence is
%and with \snc{two} sinusoidal components, 
similar to the magnetic field configurations obtained in Ref.~\onlinecite{maurer:arxiv18}, which were achieved using nanomagnet arrays compatible with current lithographic techniques. Here, we have shown that this system can support MBS even when the magnitudes of the two components of the helical field are substantially different. 

The robustness of topological superconductivity to ellipticity of the helical magnetic field depends strongly
on the magnitude of the dominant field component.
If the magnitude of this component is optimized, by making its value twice the superconducting pairing energy $\Delta$, then a topological phase appears, even when the other magnetic field component is small.
Analytic theory for an infinite length wire with a linearized electronic spectrum provides an excellent guide for interpretting results obtained numerically for a discretized model using finite-length wires.

We have also investigated the localization length of the MBS. The dependence of the energy on wire length, for the lowest energy excitation, is consistent with a simple picture in which the energy is proportional to the overlap of two exponentially localized states at the ends of the wire.
%\snc{In the limit of weak superconducting pairing $\Delta$, MBS are supported only when the chemical potential is tuned so that the one band of electrons is ungapped and the other is not, so that the Fermi energy crosses one electron band. }
%\snc{Calculations performed at different values of the chemical potential can all be interpreted in terms of this
%simple picture.}
%a larger magnetic field by itself does not guarantee the appearance of MBS fermions, as asymmetry produces an additional gap that can eliminate the necessary states. 
%This feature, however, can be improved by tuning the chemical potential closer to the gap edge. 
%To take into account the imperfections of the system, e.g., nanomagnet shape imperfections, w
%\snc{We also consider magnetic field configurations with higher harmonics and find that they do not affect the MBS
%even when they have rather large amplitudes.  These results are also consistent with the conclusion that MBS depend
%strongly on whether or not there is one ungapped band at the Fermi level.}
% considered higher order harmonics of the magnetic field. 
%We showed that there is a limit after which even higher order harmonics do not affect MBS even if they have rather large amplitude. We also showed that MBS are rather robust to low numbers of higher harmonics too, provided their amplitude is small with respect to the main harmonic. This means that some regular deviations of the magnetic field components from the sinusoidal shape we assume are not crucial for the MBS presence. 

Our results provide evidence that using lithographically  patterned micromagnets is a viable method for creating  spatially-dependent magnetic fields that, together with proximity-induced superconductivity, can be used to generate MBS.  Because intrinsic spin-orbit coupling is not required, many materials systems could also be suitable hosts for MBS, in addition to the silicon wires considered here.

\begin{acknowledgments} 
We thank Anton Akhmerov, Ryan Foote, Alex Levchenko, Constantin Schrade, Brandur Thorgrimsson, and Hongyi Xie for helpful discussions. This work was supported by the Vannevar Bush Faculty Fellowship program sponsored by the Basic Research Office of the Assistant Secretary of Defense for Research and Engineering and funded by the Office of Naval Research through Grant No. N00014-15-1-0029, by NSF EAGER Grant No. DMR-1743986, by the Army Research Office, Laboratories for Physical Sciences Grant No. W911NF-18-1-0115. The views and conclusions contained here are those of the authors and should not be interpreted as representing the official policies, either expressed or implied, of the Office of Naval Research or the U.S. Government. 
\end{acknowledgments}

\end{document}